\documentclass[aps,prb,twocolumn,showpacs,showkeys]{revtex4}

\usepackage{graphicx,color,subfig,amsmath,bm,amssymb}
\usepackage[USenglish]{babel}

\begin{document}

\title{Tight-binding models for the new iron based superconductor materials}

\author{Helmut Eschrig}
\author{Klaus Koepernik}

\email{h.eschrig@ifw-dresden.de}
\homepage{http://www.ifw-dresden.de/~/helmut}

\affiliation{IFW Dresden, PO Box 270116, D-0111171 Dresden, Germany}

\begin{abstract}
  The rich novel materials class of iron based superconductors turned
  out to exhibit a very complex electronic structure, despite of the
  simplicity of their crystal structures. For various approaches to
  study the instability against magnetic order or superconductivity, a
  real space description of the electronic structure is required. Here,
  the bonding situation and the orbital structure of the electronic
  state are analyzed and minimum tight-binding models quantitatively
  correctly describing the low-energy electronic structure are provided.
\end{abstract}


\maketitle

\section{Introduction}

The discovery of high-temperature superconductivity in La(O,F)FeAs\ \
\cite{r01} immediately raised great expectations to advance the theory
of high-temperature superconductivity in general, since the crystal
structure appeared quite simple and on a first glance displayed
similarity to that of the cuprates. It quickly turned out, however, that
there are also striking differences, in particular of the correlated
nature of the electron state. The analysis of many problems in this
respect demands real space approaches based on tight-binding (tb)
models. An early paper in this direction used a two-orbital tb
model.\cite{r02} The need for gradually improving such models was
already clearly seen in this paper. Unfortunately, improvements turned
out to grow terribly in complexity, and hopping matrix elements up to
the fifth neighbor matter.\cite{r0} The situation clearly calls for a
detailed analysis of the electronic structure in the tb language as
complete as possible. The present paper tries to contribute to this
task.

Among the new iron based superconducting materials there are essentially
four structural families christened 11, 111, 1111, and 122 according to
their stoichiometry in the undoped cases.  FeSe and FeTe are
representatives of the first family,\cite{r2} LiFeAs and NaFeAs
represent the second,\cite{r3}\begin{hyphenrules}{nohyphenation}
  REOFePN,\end{hyphenrules} where RE is Y, La or a light rare earth
element and PN is P, As or Sb, belong to the third,\cite{r4,r02} and
AFe$_2$As$_2$, where A is Ca, Sr or Ba, belong to the last.\cite{r5} All
families exhibit a tetragonal structure at room temperature which
slightly distorts below the N\'eel temperature of the anti-ferromagnetic
order.  The first three families have the non-symmorphic space group
P4/nmm in common, while the last family crystallizes in the symmorphic
space group I4/mmm.  The unit cells of the first three families as well
as the tetragonal cell of the last (which contains two unit cells) are
shown in Fig.~\ref{fig:0}.

\begin{widetext}
\onecolumngrid
\begin{figure}[h]
  \centering
  \includegraphics[scale=1.6,clip=]{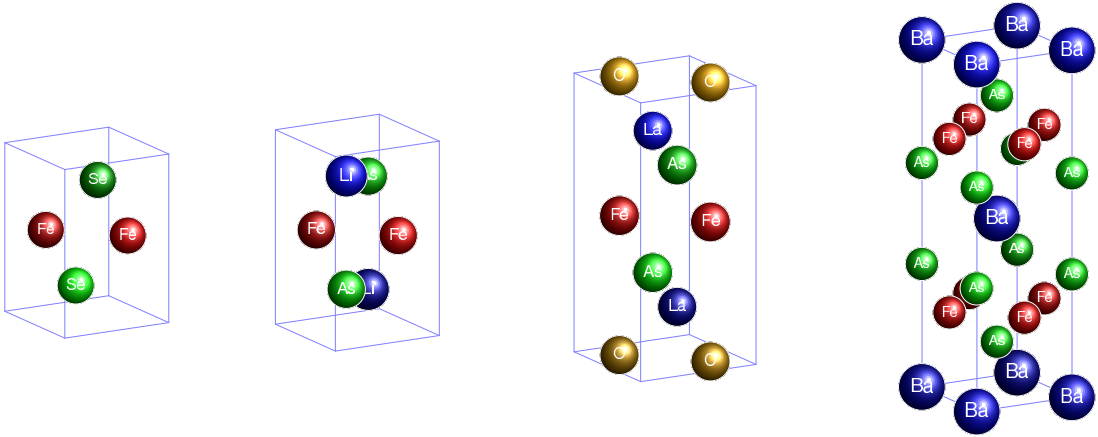}
  \caption{From left to right: unit cells of FeSe,
    LiFeAs, LaOFeAs and BaFe$_2$AS$_2$ (the tetragonal cell contains two
    unit cells in the latter case).}
  \label{fig:0}
\end{figure}
\end{widetext}

The generic structural element is a metallic iron chalcogenide/pnictide
layer consisting of a square iron atom plane sandwiched between two
chalcogen/pnictogen (c/p) atom planes in such a way that every Fe atom
is in the center of a c/p tetrahedron. The families 11 and 111 consist
of neutral metallic layers only, stacked on top of each other in the
direction of the tetragonal axis and only weakly bonded in this
direction. Thereby LiAs is isoelectronic with Se. They should easily
cleave in planes perpendicular to the tetragonal axis. In contrast, in
the 1111 and 122 families the FeAs layer is charged and anionic and is
intercalated with a non-conducting cationic layer, e.g. LaO in the 1111
family and Ba in the 122 family. Hence, these materials cannot neutrally
cleave between intact atom layers.

Phenomenologically, superconductivity seems to be quite robust in the
(doped) families 1111 and 122, but more sensitive to imperfections in
the families 11 and 111. This at least goes in line with a remarkable
structural difference revealed in Table~\ref{tab:1}. For representatives
of the four families this table shows the nearest in-plane Fe-Fe
distance $d$ in \AA, and the ratio $r$ of the distance of c/p layers
from the Fe layer related to the in-plane Fe-Fe distance.  This latter
ratio would be 0.5 in the case of regular c/p tetrahedra. As is evident
from Table~\ref{tab:1}, the cationic interlayers of the 1111 and 122
families put the Fe layer under considerable tensile strain, and in line
with that the c/p tetrahedra are considerably contracted in the
direction of the tetragonal axis. The ratio $r$ is by the way unusually
badly reproduced by lattice relaxation within non-magnetic density
functional theory (DFT), and it seems to be somewhat sensitive to
magnetic order. Which of the Fe $3d$ bands cross the Fermi level and the
sequence in energy of these bands depends on the actual value of $r$ and
often differs for the experimental value and for the DFT relaxed value.

\begin{table}[h]
  \centering
  \begin{tabular}{l|l|c|c}
    family & representative   & $d$    & $r$\\ \hline
    11     & FeSe           & 2.6653 & 0.5350 \\
    111    & LiFeAs         & 2.6809 & 0.5614 \\
    1111   & LaOFeAs        & 2.8497 & 0.4620 \\
    122    & BaFe$_2$As$_2$ & 2.8019 & 0.4855
  \end{tabular}
  \caption{Structural parameters of the four families as described in
    the text.} 
  \label{tab:1}
\end{table}

\textit{All results presented in this paper refer to the experimental
  structure data and to the non-magnetic state.}

The aim is to provide tb models of the low-energy background electronic
structure on the basis of which collective behavior like magnetism and
superconductivity may develop and may be analyzed by many-body
theories. It will be shown that the low-energy physics is essentially
more two-dimensional (2D) in the families 1111 and 122 than in the other
two, again distinguishing the lower two families of Table~\ref{tab:1}
from the upper two.

\begin{figure}[h]
  \centering
  \includegraphics[scale=0.32,angle=-90,clip=]{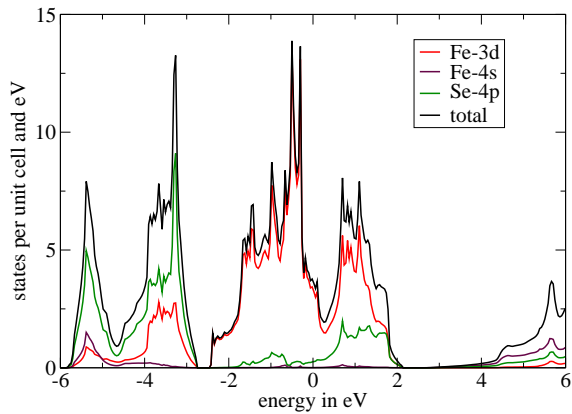}
  \caption{Partial and total DOS for FeSe. The Fermi
    level is at 0 eV.}
  \label{fig:01}
\end{figure}

On Fig.~\ref{fig:01} the essential atomic-shell resolved partial
densities of states (DOS) together with the total DOS are shown for the
experimental structure parameters of FeSe ($a=3.7734$ \AA, $c=5.5258$
\AA, $z_{Se}=0.267,\; z_{Fe}=0$).\cite{r6} All band structures in this
work are obtained with the high precision full potential local orbital
code FPLO8\cite{r7} using the Perdew-Wang 92\cite{r8} version of the
local density approximation (LDA) density functional. Generalized
gradient approximation (GGA) results would somewhat differ in the
lattice relaxation by total energy minimization. Since, however, for
given fixed structural parameters the results considered here hardly
differ at all, and all results presented in this text are for the
experimental structures, LDA was used throughout.  

For what follows it is relevant that the all electron approach FPLO8
uses a carefully optimized `chemical' basis (one basis orbital per
atomic core or valence orbital) plus at most one polarization orbital
per band.  The quality of the chemical basis results in occupation
numbers of polarization orbitals usually well below 0.02 (per spin), in
most cases even much lower (without loss in accuracy which competes with
any of the advanced all-electron full-potential tools presently
available. For instance, in a recently published work\cite{mazin08}
results for the 1111 compound LaOFeAs obtained with different band
structure methods were compared and very good aggreement between FPLO
and FPLAPW results was assessed). This is the reason why projections on
the FPLO chemical basis orbitals have a lot of chemical relevance, and
also produce nearly optimally localized Wannier orbitals from band
structures in an approach described below in Sec.~IV.

The situation shown in Fig.~\ref{fig:01} which is generic for all four
families considered here can be understood as follows: Since there are
two Fe atoms per unit cell, there is Fe-Fe homo-covalency. Put the
structural layers in the $xy$-plane stacked in $z$-direction. For zero
wave vector there are in-plane bonding combinations of Fe-$3d$ orbitals
with angular dependence $xy/r^2$ and $(x^2-y^2)/r^2$, respectively, with
the same coefficient on both sites and anti-bonding combinations with
sign-alternating coefficients. For the angular dependence $xz/r^2$ and
$yz/r^2$ it is just the other way round, while the orbitals $\sim
(z^2/r^2-3)$ form essentially non-bonding bands. Hence, there are four
Fe-Fe bonding bands, four Fe-Fe anti-bonding bands and two non-bonding
bands. The Se-$4p$ orbitals may couple with the Fe-Fe anti-bonding
combinations only, developing Fe-Se hetero-covalency with predominantly
Se-$4p$ bands below $-3$~eV and predominantly Fe-$3d$ bands above the
Fermi level (the latter put at zero band energy). The Fermi level itself
falls in a Fe-Fe covalency pseudo-gap between the bonding plus
non-bonding bands of essentially pure Fe-$3d$ character and the
anti-bonding bands of hybridized Fe-$3d$-Se-$4p$ character. This picture
prevails in large parts of the Brillouin zone (BZ) and only modifies in
the vicinity of the BZ-edges parallel to the $z$-axis (through point M
of Fig.~\ref{fig:2} below). It results in rather small hole Fermi
surfaces (FS) of Fe-Fe bonding bands around the $k_z$-axis and in
likewise small electron FS around the BZ edges parallel to this axis,
and hence in the pseudo-gap at the Fermi level. The materials are not far
from semi-metals.

\begin{figure}[h]
  \centering
  \includegraphics[scale=0.32,angle=-90,clip=]{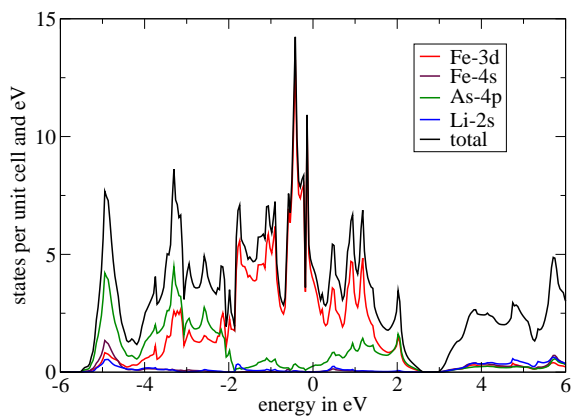}
  \caption{Partial and total DOS for LiFeAs. Experimental
  structure parameters from Ref.\cite{r9}.}
  \label{fig:02}
\end{figure}

\begin{figure}[h]
  \centering
  \includegraphics[scale=0.32,angle=-90,clip=]{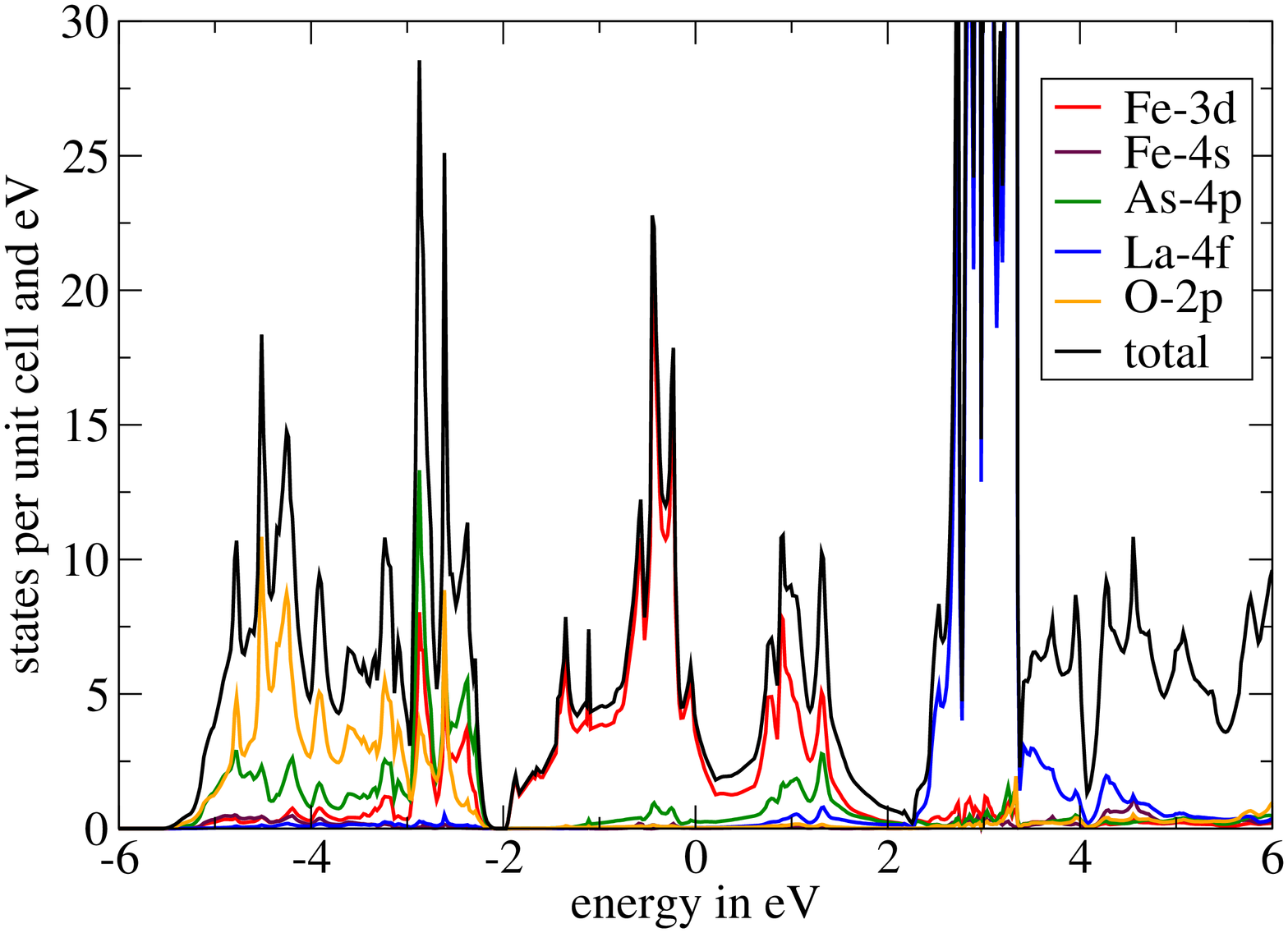}
  \caption{Partial and total DOS for LaOFeAs. Experimental
  structure parameters from Ref.\cite{r10}.}
  \label{fig:03}
\end{figure}

\begin{figure}[h]
  \centering
  \includegraphics[scale=0.32,angle=-90,clip=]{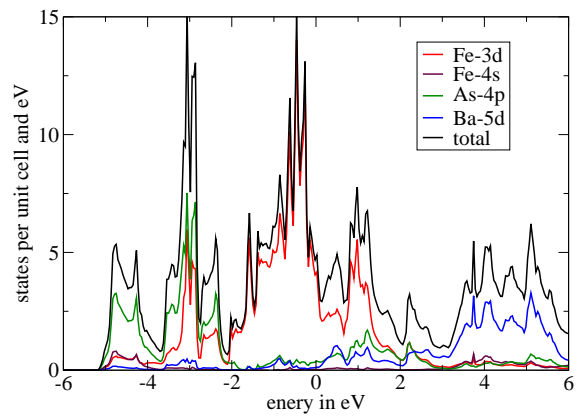}
  \caption{Partial and total DOS for BaFe$_2$As$_2$.
    Experimental structure parameters from Ref.\cite{r11}.}
  \label{fig:04}
\end{figure}

There is also Fe-$4s$-Se-$4p$ covalency with a gap of about 7~eV as seen
on the bottom of Fig.~\ref{fig:01}. The modifications in the other
families compared to 11 are gradual, see
Figs.~\ref{fig:02}--\ref{fig:04}. Since the As-$4p$ states are about 1
eV higher than the Se-$4p$ states, the gap below $-2$~eV is less
pronounced or absent in the arsenides.  In LiFeAs the pseudo-gap at
the Fermi level is less pronounced too as the Fermi radii are somewhat
larger (larger energy overlap between Fe-Fe bonding and anti-bonding
band groups).  In LaOFeAs and BaFe$_2$As$_2$ there is also no gap above
2~eV as the La-$4f$ states and the Ba-$5d$ states, respectively, come
down close to the Fermi level. (About 0.4 Ba-$5d$ states are occupied in
BaFe$_2$As$_2$ while, when doping with K, the K-$3d$ states remain
empty; this is why one K atom when replacing Ba donates only about 0.6
holes, or 0.3 holes per Fe atom, into the Fe-$3d$ bands.)  Moreover, the
O-$2p$ states of LaOFeAs appear below $-2$~eV in addition to the As-$4p$
states.  In the range between $-2$~eV and the Fermi level there is not so
much difference in the bands being almost purely Fe-$3d$ in all cases.

\begin{figure}[h]
  \centering
  \includegraphics[scale=0.32,angle=-90,clip=]{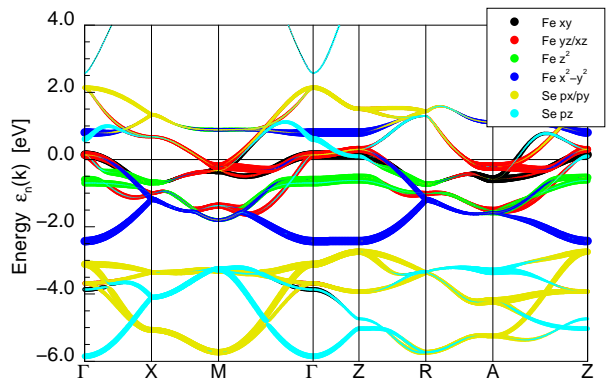}
  \caption{Orbital projection of the LDA band structure of FeSe.
(See Fig.~\ref{fig:2} for the points $\Gamma$, $X$, $M$; $Z$, $R$, $A$
are above the former at $k_z=\pi$.)}
\label{fig:05}
\end{figure}

The whole situation with Fe-$3d$-Fe-$3d$ and Fe-$3d$-c/p-$4s$ covalency
is illustrated for the generic example of FeSe in Fig.~\ref{fig:05}. The
projection of the LDA bands on the FPLO chemical basis orbitals is given
as line thickness in color code. (The diameter of the color balls in the
legend of the figure indicates the thickness for 100 percent orbital
content in the band wave function.) The cyan (c/p-$pz$) band on the line
$\Gamma$-Z which disperses down close to the Fermi level has a black
component (Fe-$xy$) hidden behind the cyan and amounting to about half
of the band state near Z. The narrowness below the Fermi level of the
non-bonding Fe-$z^2$ bands is clearly seen.  Again seen is that the
Fe-Fe bonding bands below the Fermi level do not hybridize with the c/p-$4p$
orbitals while Fe-Fe anti-bonding bands strongly hybridize. Due to the
far extend of the radial orbitals, the c/p-$4p$ states extend even above
the Fe-$3d$ bands, over an energy range of more than 10 eV. It is also
seen that the Fermi level is crossed by Fe $xy$, $xz$ and $yz$ bands
only with a small admixture of c/p $4pz$ orbitals. Right below Fermi
level Fe $x^2-y^2$ and $z^2$ admixture sets in, so that one fails to
reproduce Fermi velocities even qualitatively without consideration of
the latter orbitals and their corresponding bands. 

Remarkable in this respect is also the orbital anisotropy of the FS,
again in stark contrast to the case of the cuprates. While the three FS
hole cylinders around $\Gamma$-Z appear more or less orbitally isotropic
with one of the outer cylinders and the inner cylinder of $xz-yz$
character (of course with their mixing rotating together with $\bm k$)
and the other outer cylinder of $xy$ character, the outer FS electron
cylinder around M in the $k_z=0$ plane appears totally orbitally
anisotropic with $xz-yz$ character in the directions towards X and
nearly pure $xy$ character in the directions towards $\Gamma$. The inner
electron cylinder is again more or less orbitally isotropic with $xz-yz$
character. In the $k_z=\pi$ plane the inner and outer electron cylinders
have interchanged their orbital character.

\begin{figure}[h]
  \centering
  \includegraphics[scale=0.32,angle=-90,clip=]{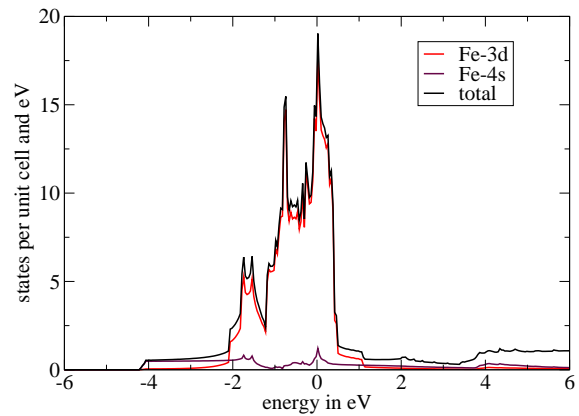}
  \caption{Partial and total DOS for a hypothetical
    structure obtained by removing the Se atoms from FeSe.}
  \label{fig:06}
\end{figure}

In all those considerations the presence of the c/p atoms is essential.
Without there presence, the unit cell would reduce to containing one Fe
atom only, and no interband Fe-Fe covalency could develop. In
Fig.~\ref{fig:06} the DOS of a hypothetical lattice is shown with the Se
atoms removed from the FeSe crystal, but the Fe atoms left at their
positions. The Fe-$3d$ occupation remains nearly the same (Mullikan
analysis with the FPLO8 basis results in charge transfer between Fe and
c/p of less than 0.1 electron charge/atom), but no pseudo-gap whatsoever
appears as there are now only five Fe-$3d$ bands in the doubled BZ.
There is also no covalency band gap of the Fe-$4s$ states any more.
Fe-Fe covalency appears only due to the doubling of the unit cell caused
by the presence of the c/p atoms, due to their additional potential and
due to hybridization with c/p-$p$ orbitals. Also, the width and depth of
the pseudo-gap are quite sensitive to the c/p position, that is, to the
actual value of the Wyckoff parameter $z_{c/p}$.

In a vicinity of half an eV around the Fermi level, the band structure
in the 1111 family exhibits a pronounced quasi-2D character. It is
somewhat less 2D and in a narrower energy window at the Fermi level in the
122 family and much more 3D in the 11 and 111 families.

In the next section the implications of symmetry on the electronic
structure are analyzed in detail. This provides some key to the
construction of reduced tb models. In Sec.~III, a minimum 2D tb model is
derived for reasonable quantitative approximations of the low-energy
band structures of the Fe-c/p layers (the focus being essentially on
excitation energies below 0.1~eV), and parameters are given for
representatives of all four families. These are based on a Wannier
function representation introduced in Sec.~IV. The families 11 and 111
exhibit quite sizable 3D dispersion close to the Fermi level.
Corresponding 3D tb models for these cases are given in Sec.~V, and
finally the results are summarized in Sec.~VI.

\section{Symmetry of the iron chalcogenide/pnictide layer}

The 11 family consists of the Fe-c/p layers only, generic for all four
families. To be specific, FeSe is taken as example in what follows. We
start with a more detailed description of the FeSe structure:
\begin{figure}[h]
  \centering
  \includegraphics[scale=1]{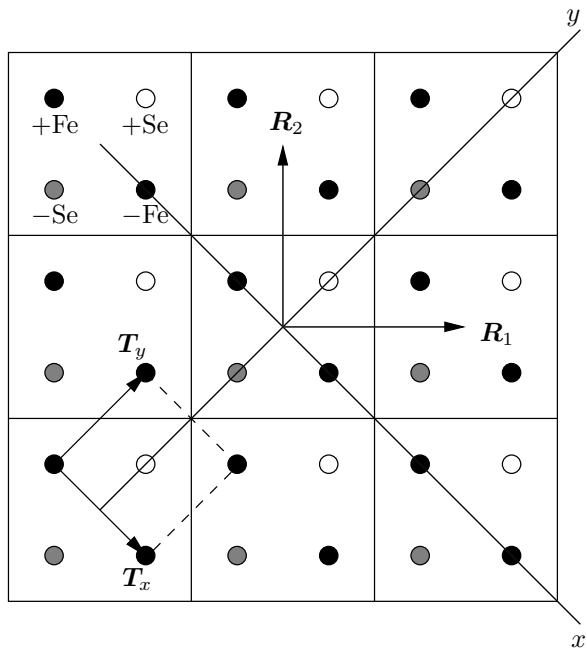}
  \caption{Tetragonal plane of nine unit cells (full-line quares)
    spanned by lattice basis vectors $\bm R_1$ and $\bm R_2$ of FeSe.
    The open circles mark Se positions above and the shadowed circles
    below the drawing plane. The Fe sites (full circles) are located in
    the drawing plane. If the Se sites are neglected, the half-size unit
    cell is spanned by $\bm T_x$ and $\bm T_y$ as shown in the left
    lower corner.}
  \label{fig:1}
\end{figure}

The two Fe and Se positions per unit cell are symbolically distinguished
by a sign. Atom positions are (Fig.~\ref{fig:1})
\begin{equation}
  \label{eq:01}
  \begin{split}
  \bm S_{\pm Fe} &= \frac{1}{4}\bigl(\mp\bm R_1 \pm\bm R_2\bigr),\\
  \bm S_{\pm Se} &= \frac{1}{4}\bigl(\mp\bm R_1 \mp\bm R_2\bigr)
                                     \pm z_{Se}\bm R_3.
  \end{split}
\end{equation}
The vectors $\bm T_x$ and $\bm T_y$ of Fig.~\ref{fig:1} are not lattice
vectors of the full structure. However, in combination with a reflection
$n$ on the $(x,y)$-plane they nevertheless form non-symmorphic symmetry
elements $(n|\bm T_x)$, $(n|\bm T_y)$ (diagonal glide plane, $(n|\bm
T_x)\phi(\bm r) = \phi(n^{-1}\bm r-\bm T_x),\; n^{-1}=n$). Applied to a
Bloch state $\psi_{\bm k\nu}(\bm r)$, any space group element yields
$(A|\bm T_A+\bm R_l)\psi_{\bm k\nu}(\bm r) = \exp(i(A\bm k)\cdot\bm
R_l)\psi^\prime_{A\bm k}(\bm r)$, where $\nu$ is the band index, and
$\psi^\prime_{A\bm k}(\bm r)$ is a linear combination of states
$\psi_{A\bm k\nu^\prime}(\bm r)$
 with the restriction  $\varepsilon_{A\bm k\nu^\prime} =
\varepsilon_{\bm k\nu}$ on the band energies. ($\bm T_n =
\bm T_x$, some one-one relation $A\leftrightarrow \bm T_A$ must be
chosen, $(n|\bm T_y) = (n|\bm T_n + \bm R_2)$. Moreover, the result of
$(A|\bm T_A+\bm R_l)\psi_{\bm k\nu}(\bm r)$ is based on a chosen phase
relation between $\psi_{\bm k\nu}(\bm r)$ and $\psi^\prime_{A\bm k}(\bm
r)$, which choice is made here for the sake of simplicity different from
that later in (\ref{eq:08a}) and (\ref{eq:14}); the results considered
here do not depend on this choice.) One has $(A|\bm T_A)(B|\bm T_B) =
(AB|\bm T_{AB} + \bm R_{AB})$, where $\bm R_{AB}$ is a lattice vector.
If, for the wave vector group of $\bm k$, $\{(A|\bm T_A + \bm
R_l)\,|\,A\bm k = \bm k + \bm G_A\}$ where $\bm G_A$ is a reciprocal
lattice vector, one defines the multiplier representation
$\rho^A_{\nu'\nu}(\bm k) = \exp(-i\bm k\cdot\bm R_l)\langle\psi_{\bm
  k\nu'}| (A|\bm T_A + \bm R_l)|\psi_{\bm k\nu}\rangle$, then
\begin{equation}
  \label{eq:02}
    \rho^A \rho^B = \exp(i\bm k\cdot\bm R_{AB})\; \rho^{AB}
\end{equation}
follows. Time inversion $T$ is defined as the anti-unitary operator
$T\phi(\bm r) = \phi^*(\bm r)$ and hence $T\psi_{\bm k\nu} = \psi_{-\bm
  k T\nu}$ and $T(A|\bm T_A + \bm R_l) = (A|\bm T_A + \bm R_l)T$. Now,
if $m_2$ is the reflection plane through iron sites sending $\bm R_2
\rightarrow -\bm R_2$, then the anti-unitary operator $\Theta =
Tm_2(n|\bm T_n)$ is an element of the wave vector group for any wave
vector $\bm k$ with $\bm k\cdot\bm R_1 = \pm\pi$, that is, on the face
of the Brillouin zone (BZ) with central point $X$ (Fig.~\ref{fig:2}
below), and $\rho^\Theta(\bm k)\,\rho^\Theta(\bm k) = -1$ for any such
$\bm k$ since $\Theta^2 = (1|\bm R_1)$. The existence of an anti-unitary
operator with square equal to $-1$ which commutes with the Hamiltonian
is an analogue to Kramers degeneracy,\cite{r1} and an immediate
consequence here is a twofold degeneracy of all bands on the four side
faces of the BZ. ($\bm R_1$ and $\bm R_2$ may be interchanged in the
above considerations, with $m_2$ replaced with $m_1$.) One also has
$-(m_1(n|\bm T_n) + (n|\bm T_n)m_1)/2 = 2^2_1,\; (m_2(n|\bm T_n) +
(n|\bm T_n)m_2)/2 = 2^1_1$ where $2^2_1$ is a screw rotation by $\pi$
around the axis through the lattice vector $\bm R_2$ of
Fig.~\ref{fig:1}, accompanied by a glide by the vector $\bm R_2/2$, and
$2^1_1$ is the corresponding screw rotation around $\bm R_1$ as given in
that figure.  The notation of space group elements follows
Hermann-Mauguin\cite{r12}.  We prefer to use the two reflections in the
considerations of hopping matrix elements below. 

This is all generally
true for the P4/nmm space group (which is also the space group of a
single FeSe layer of Fig.~\ref{fig:1}). Hence, it holds for the 11, 111
and 1111 materials (FeSe, LiFeAs and LaOFeAs as examples).  It does not
hold for the 122 materials (e.g. BaFe$_2$As$_2$) which crystallize in
the I4/mmm structure (body-centered tetragonal).  The space group I4/mmm
is symmorphic, and hence there appear no multipliers different from
unity in representations of wave vector groups and no two-fold
degeneracies on whole BZ faces. These degeneracies only reappear, if one
neglects $z$-axis dispersion. In the following we first focus on the
layer structure of Fig.~\ref{fig:1} which is representative for all four
families, if $z$-axis dispersion is neglected.

The full crystal potential $V$ must be invariant under all symmetry
operations including lattice translations and glide reflections. With
respect to the latter, it can be decomposed into a part $V_+$ that is
invariant under both the $\bm T_x$- or $\bm T_y$-translations and the
reflection $n$ on the $(x,y)$-plane separately and a part $V_-$ that is
alternating with respect to both separate operations. This potential
representation will be used in the next section. 

Further important symmetry transformations of all structures with the
P4/nmm space group are a $\pi$-rotation $2^x$ around the $x$-axis of
Fig.~\ref{fig:1}, a $\pi$-rotation $2^y$ around a parallel to the
$y$-axis through Fe sites and the spatial inversion $\bar 1$ with its
center halfway between any of the pairs $\pm X$ of atoms (all pairs have
the same midpoint). Non-zero Hamiltonian matrix elements must either be
invariant under symmetry transformations of the structure or transform
into each other.

Due to the presence of the glide plane $(n|\bm T_n)$, all atom positions
within the unit cell of the layer of Fig.~\ref{fig:1} or more generally
of a crystal with P4/nmm space group come in symmetry equivalent pairs
$\pm X$ like $\pm$Fe and $\pm$Se in Fig.~\ref{fig:1}. Denote local site
orbitals of a local orbital basis correspondingly by a sign superscript
as $\phi^\pm$. (This superscript is to be well distinguished from the
sign subscript on $V_\pm$ for the component of the crystal potential
which has a different meaning defined above, although related to the
same symmetry element.) Expanded in this basis, the Hamiltonian may be
brought into a block matrix structure
\begin{equation}
  \label{eq:03}
  H =
  \begin{pmatrix}
    H^{++} & H^{+-} \\ H^{-+} & H^{--}
  \end{pmatrix}\,,
\end{equation}
where, if one denotes hopping matrix elements by $t^{rst}_{ij}$ for
hopping a distance $r \bm T_x + s \bm T_y + t \bm R_3,\; r,s$ integer,
between local site orbitals $\phi_i$ and $\phi_j$ (of either
superscript), then hoppings with $r+s$ even (accompanied with
corresponding Bloch phase factors) enter $H^{++}$ and $H^{--}$
while hoppings with $r+s$ odd enter $H^{+-}$ and $H^{-+}$.

Note that the anti-unitary transformation $T\bar 1$ is an element of the
wave vector group for every wave vector $\bm k$, this time with $(\rho^{T\bar
1})^2 = +1$ and hence not causing a degeneracy. The local site orbitals
may always be chosen so that
\begin{equation}
  \label{eq:04a}
  T\bar 1\phi^\pm = \phi^\mp,
\end{equation}
this time also for the 122 structure,
in all cases with respect to the center of inversion $\bar 1$ half-way
between nearest Fe neighbors.

For instance, if in a natural way the local site orbitals are parity
eigenstates, appropriate imaginary factors at odd parity orbitals do.
Then, for the hopping matrix elements one has $\langle\phi^-|\hat
H|\phi^-\rangle = \langle T\bar 1\phi^-|\hat H|T\bar 1\phi^-\rangle^* =
\langle\phi^+|\hat H|\phi^+\rangle^*$, and $\langle\phi^-|\hat
H|\phi^+\rangle = \langle\phi^+|\hat H|\phi^-\rangle^*$. It immediately
follows that with this basis choice (which is always possible and even
the natural one)
\begin{equation}
  \label{eq:04}
  H =
  \begin{pmatrix}
    H^{++} & H^{+-} \\ {H^{+-}}^* & {H^{++}}^*
  \end{pmatrix}\,.
\end{equation}
Since generally $H^{-+} = {H^{+-}}^\dag$, it also follows that the block
matrix $H^{+-}$ must be symmetric:
\begin{equation}
  \label{eq:05}
  {H^{+-}}^t = H^{+-},
\end{equation}
where $H^t$ means the transposed of $H$. This reduces the calculation of
Hamiltonian (and overlap) matrix elements by a factor of two.

\section{Tight-binding bands for the iron chalcogenide/pnictide layer}

In this section tb models for the iron c/p layers are considered. Again,
to be specific, the discussion is first for FeSe and then generalized.

Fe $3d$-orbitals are denoted by their angular character and by an upper
sign indicating centering at site $\bm S_{\pm Fe}$ as in the previous
section for the general case. For instance $(ixz)^+$ means a real
orbital with angular dependence $xz/r^2$ centered at $\bm S_{+Fe}$,
multiplied with the imaginary unit factor later being used to get real
Hamiltonian matrices; $(z^2)$ abbreviates the real orbital with angular
dependence $(3z^2-r^2)/r^2$.  The following site orbitals $\phi^\pm$
within a unit cell are introduced which obey (\ref{eq:04a}):
\begin{equation}
  \label{eq:06}
  \begin{array}{rcrc}
    1:&\;(xy)^+,     \quad & 6:&\;(xy)^-,\\
    2:&\;(x^2-y^2)^+,\quad & 7:&\;(x^2-y^2)^-,\\
    3:&\;(ixz)^+,    \quad & 8:&\;(-ixz)^-,\\
    4:&\;(iyz)^+,    \quad & 9:&\;(-iyz)^-.\\
    5:&\;(z^2)^+,    \quad &10:&\;(z^2)^-.
  \end{array}
\end{equation}
The numbering of orbitals will later be used for (lower) matrix indices
of the hopping matrices $t_{ij}$.  If in what follows no superscript is
applied, then $\phi$ means an orbital either on site $\bm S_{+Fe}$ or on
$\bm S_{-Fe}$.

There are 4 distinct onsite Hamiltonian diagonal matrix elements (the subscript
$i$ being again the number from (\ref{eq:06}))
\begin{equation}
  \label{eq:07}
  \epsilon_i, \quad \epsilon_4 = \epsilon_3, \quad
  \epsilon_{i+5} = \epsilon_i\,.
\end{equation}
while due to orbital onside orthogonality there are no onside
Hamiltonian off-diagonal matrix elements.  The $10 \times 10$
Hamiltonian matrix in site orbital representation has an obvious $5
\times 5$ block structure (\ref{eq:04}, \ref{eq:05}).

Since only one layer is considered, only in-plane hoppings $t^{rs}_{ij}$
a distance $r\bm T_x + s\bm T_y$ figure, a fact relevant in what
follows.  Now one has to distinguish matrix elements, the orbital
products of which are even with respect to reflection on the
$(x,y)$-plane and which hence are coupled by the kinetic energy operator
and by $V_+$, from the ones, the orbital products of which are odd with
respect to reflection on the $(x,y)$-plane and which hence are coupled
by $V_-$.  The former are (superscripts dropped, $i,j = 1,\ldots,5$)
$t_{ii}$, $t_{12}$, $t_{15}$, $t_{25}$, $t_{34}$, their transposed and
the corresponding $t_{i+5,j+5}$, $t_{i,j+5}$ and $t_{i+5,j}$.  The later
are $t_{ij},\; i=1,2,5;\, j=3,4$ their transposed and the corresponding
$t_{i+5,j+5}$, $t_{i,j+5}$ and $t_{i+5,j}$.

For the former one has immediately $t_{i+5,j+5} = t_{ij}$ and $t_{i+5,j}
= t_{i,j+5}$, respectively, while for the latter the same relations are
the consequence of our phase choices in (\ref{eq:06}). With our choice
of the orbitals the former matrix elements are real while the latter are
purely imaginary. The matrix elements $t_{55}$ and $t_{5,10}$ between
the essentially non-bonding orbitals are small and are neglected further
on.

Considering the symmetry transformations of the structure it is easily
found that the 9 distinct non-zero first neighbor hoppings are
$t^{10}_{16} = t^{\bar 10}_{16} = t^{01}_{16} = t^{0\bar 1}_{16}$,
$t^{10}_{18} = -t^{\bar 10}_{18} = t^{01}_{19} = -t^{0\bar 1}_{19}$,
$t^{10}_{27} = \cdots$, $t^{10}_{29} = -t^{01}_{28} = \cdots$,
$t^{10}_{2,10} = \cdots$, $t^{10}_{38} = t^{01}_{49} = \cdots$,
$t^{10}_{49} = t^{01}_{38} = \cdots$, and $t^{10}_{4,10} = t^{01}_{3,10}
= \cdots$, where we did not spell out all relations which follow rather
obviously from the above mentioned symmetry transformations of the
structure and of the orbital products from which also the relations
$t^{mn}_{i,j+5} = t^{mn}_{j,i+5},\; m+n$ odd follow. As an example, one
of the above relations, $t^{10}_{18} = -t^{\bar 10}_{18}$, is obtained
by a $\pi$-rotation $2^y$ which transforms $(xy)^+$ into $-(xy)^+$ and
$(ixz)^-$ into itself shifted by a translation $(m,n) = (-2,0)$.
$t^{10}_{18} = t^{01}_{19}$ is obtained by an $m_2$-reflection. As an
example of the last mentioned type ($t^{mn}_{i,j+5} = t^{mn}_{j,i+5}$),
$t^{10}_{18} = t^{10}_{36}$ is obtained by a $\pi$-rotation $2^x$
changing the sign of the orbital $(xy)^+$ but not that of $(-ixz)^-$,
and a subsequent shift by $\bm T_x$ mapping $(xy)^+$ to $(xy)^-$ and
$(-ixz)^-$ to $((ixz)^+)^*$. Since $(ixz)^+$ couples to $(xy)^-$ via
$V_-$, the $\bm T_x$-shift changes the sign of the matrix element back.
On the other hand, in $t^{10}_{2,10} = t^{10}_{57}$ no sign change of
orbitals happens in the $\pi$-rotation $2^y$, and no sign change in the
shift by $\bm T_x$ since this time the orbitals couple via the kinetic
energy operator and $V_+$.

There are also 9 distinct non-zero second neighbor hoppings
$t^{11}_{ij},\, i,j=1,\ldots,5$, while $t^{11}_{12} = 0$, $t^{11}_{25} =
0$, $t^{11}_{14} = t^{11}_{13}$, $t^{11}_{24} = -t^{11}_{23}$, both
latter cases by an $m_2$-reflection, $t^{11}_{44} = t^{11}_{33}$,
$t^{11}_{45} = t^{11}_{35}$ and $t^{mn}_{ij} = t^{mn}_{ji},\; m+n$ even.
For instance for $ij=11$ the latter relations are obtained by an
$m_1$-reflection, and, for $i$ or $j$ or both equal to 3 or 4, by an
additional $m_2$-reflection. As already obtained in the general case,
$t^{11}_{i+5,j+5} = t^{11}_{ij}$.

Since we try to limit the number of parameters as much as possible, and
we focus on the low-energy vicinity of the Fermi level, we neglect
$t_{55}$ and $t_{5,10}$ for the Fe-Fe non-bonding bands which stay away
from the Fermi level by more than half an eV. (Note, however, that for the
LDA or GGA \textit{relaxed} structure parameters these bands may even
cross the Fermi level; so far there are nevertheless no reasons to
assume that this happens in nature.) The hybridization parameters
$t_{i5},\; i=1,\ldots,4$ however cannot be neglected. They strongly
influence some Fermi velocities.

The hoppings $t^{11}$ happen essentially through the c/p site. Due to
the large radial extent of the c/p-$4p$ orbitals, nearest neighbor
c/p-c/p hopping along the c/p tetrahedron edges is strong, and hence the
effective Fe-Fe hoppings $t^{20},\; t^{21}$ and $t^{22}$ are not yet
small. In order to fit besides the FS radii also the Fermi velocities
reasonably well, they must be taken into account at least for the
orbitals 1,3,4 and 6,8,9.

\begin{figure}[h]
  \centering
  \includegraphics[scale=1]{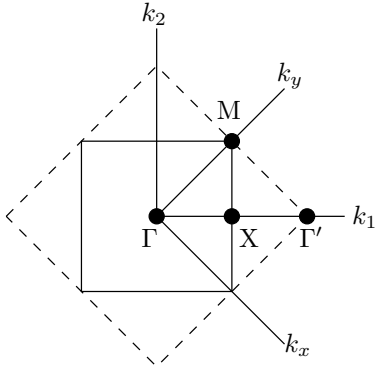}
  \caption{BZ of the structure of Fig.~\ref{fig:1}, dashed for the Fe
    structure alone (one Fe atom per cell) and full lines for the full
    FeSe structure. The wave number components are scaled so that the
    symmetry points are $(k_x,k_y) = (0,0)$ for $\Gamma$,
    $(\pi/2,\pi/2)$ for X, $(\pi,0)$ for M and $(\pi,\pi)$ for
    $\Gamma'$, as well as $(k_1,k_2) = (0,0)$ for $\Gamma$, $(\pi,0)$
    for X, $(\pi,\pi)$ for M and $(2\pi,0)$ for $\Gamma'$.}
  \label{fig:2}
\end{figure}

Fig.~\ref{fig:2} shows the two-dimensional BZ of the FeSe slab of
Fig.~\ref{fig:1}. A notation
\begin{equation}
  \label{eq:08}
  k_1 = k_x + k_y, \quad k_2 = -k_x + k_y
\end{equation}
is used, and Bloch sums are defined as
\begin{equation}
  \label{eq:08a}
  \phi_{\bm  ki} \sim \sum_l \phi_i e^{i\bm k\cdot(\bm R_l + \bm S_i)},
\end{equation}
where $\bm S_i$ is the site of the basis orbital $\phi_i$.

The Hamiltonian matrix $H^{++}$ in Bloch state representation is
\begin{equation}
  \label{eq:09}
  \begin{split}
    H^{++}_{11} &= \epsilon_1 + 2t^{11}_{11}(\cos k_1 + \cos k_2)\;+\\
                &\quad + 2t^{20}_{11}(\cos(2k_1) + \cos(2k_2)),\\
    H^{++}_{12} &= 0,\\
    H^{++}_{13} &= 2it^{11}_{13}(\sin k_1 - \sin k_2),\\
    H^{++}_{14} &= 2it^{11}_{13}(\sin k_1 + \sin k_2),\\
    H^{++}_{15} &= 2t^{11}_{15}(\cos k_1 - \cos k_2),\\
    H^{++}_{22} &= \epsilon_2 + 2t^{11}_{22}(\cos k_1 + \cos k_2),\\
    H^{++}_{23} &= 2it^{11}_{23}(\sin k_1 + \sin k_2),\\
    H^{++}_{24} &= 2it^{11}_{23}(-\sin k_1 + \sin k_2),\\
    H^{++}_{25} &= 0,\\
    H^{++}_{33} &= \epsilon_3 + 2t^{11}_{33}(\cos k_1 + \cos k_2)\;+\\
                &\quad + 2t^{20}_{33}\cos(2k_x) 
                   + 2t^{02}_{33}\cos(2k_y)\;+\\
                &\quad + 4t^{22}_{33}\cos(2k_x)\cos(2k_y),\\
    H^{++}_{34} &= 2t^{11}_{34}(\cos k_1 - \cos k_2),\\
    H^{++}_{35} &= 2it^{11}_{35}(\sin k_1 + \sin k_2),\\
    H^{++}_{44} &= \epsilon_3 + 2t^{11}_{33}(\cos k_1 + \cos k_2)\;+\\
                &\quad + 2t^{02}_{33}\cos(2k_x) 
                   + 2t^{20}_{33}\cos(2k_y)\;+\\
                &\quad + 4t^{22}_{33}\cos(2k_x)\cos(2k_y),\\
    H^{++}_{45} &= 2it^{11}_{35}(\sin k_1 - \sin k_2),\\
    H^{++}_{55} &= \epsilon_5,\\
    H^{++}_{ji} &= (H^{++}_{ij})^*.
  \end{split}
\end{equation}
Of course, $H^{++}$ is Hermitian, and in agreement with Hermiticity and
our phase choice in (\ref{eq:02}) one finds that $t^{11}_{ij}$ is real for
$t^{11}_{ij} = t^{11}_{ji}$ and imaginary for $t^{11}_{ij} = -t^{11}_{ji}$. Hence, one
finds that $H^{++}_{ij}$ is real. Together with (\ref{eq:04}) this also
implies
\begin{equation}
  \label{eq:10}
  H^{--} = H^{++}.
\end{equation}
This reality is the consequence of absence of hoppings in $\bm R_3$
direction and does not hold in the more general case.

For $H^{+-}$ one finds
\begin{equation}
  \label{eq:11}
  \begin{split}
     H^{+-}_{16}   &= 2t^{10}_{16}(\cos k_x + \cos k_y)\;+\\
                   &\quad + 2t^{21}_{16}((\cos k_1 + \cos k_2)
                                         (\cos k_x + \cos k_y)\;-\\
                   &\qquad\qquad\quad - \sin k_1(\sin k_x + \sin k_y)\;+\\
                   &\qquad\qquad\quad + \sin k_2(\sin k_x - \sin k_y)),\\
     H^{+-}_{17}   &= 0,\\
     H^{+-}_{18}   &= 2it^{10}_{18}\sin k_x,\\
     H^{+-}_{19}   &= 2it^{10}_{18}\sin k_y,\\
     H^{+-}_{1,10} &= 0,\\
     H^{+-}_{27}   &= 2t^{10}_{27}(\cos k_x + \cos k_y),\\
     H^{+-}_{28}   &= -2it^{10}_{29}\sin k_y,\\
     H^{+-}_{29}   &= 2it^{10}_{29}\sin k_x,\\
     H^{+-}_{2,10} &= 2t^{10}_{2,10}(\cos k_x - \cos k_y),\\
     H^{+-}_{38}   &= 2t^{10}_{38}\cos k_x + 2t^{10}_{49}\cos k_y\;+\\
                   &\quad + 2t^{21}_{38}((\cos k_1 + \cos k_2)\cos k_x\;-\\
                   &\qquad\qquad\quad -(\sin k_1 - \sin k_2)\sin k_x)\;+\\
                   &\quad + 2t^{21}_{49}((\cos k_1 + \cos k_2)\cos k_y\;-\\
                   &\qquad\qquad\quad -(\sin k_1 + \sin k_2)\sin k_y),\\
     H^{+-}_{39}   &= 0,\\
     H^{+-}_{3,10} &= 2it^{10}_{4,10}\sin k_y,\\
     H^{+-}_{49}   &= 2t^{10}_{49}\cos k_x + 2t^{10}_{38}\cos k_y\;+\\
                   &\quad + 2t^{21}_{49}((\cos k_1 + \cos k_2)\cos k_x\;-\\
                   &\qquad\qquad\quad -(\sin k_1 - \sin k_2)\sin k_x)\;+\\
                   &\quad + 2t^{21}_{38}((\cos k_1 + \cos k_2)\cos k_y\;-\\
                   &\qquad\qquad\quad -(\sin k_1 + \sin k_2)\sin k_y),\\
     H^{+-}_{4,10} &= 2it^{10}_{4,10}\sin k_x,\\
     H^{+-}_{5,10} &= 0.
  \end{split}
\end{equation}
$H^{+-}$ is again real, and hence
\begin{equation}
  \label{eq:12}
  H^{+-} = H^{-+}.
\end{equation}
In total, the block structure (\ref{eq:04}) specializes to
\begin{equation}
  \label{eq:13}
  H =
  \begin{pmatrix}
    H^{++} & H^{+-} \\ H^{+-} & H^{++}
  \end{pmatrix}, \quad (H^{+-})^t = H^{+-}.
\end{equation}

This Hamiltonian is easily block diagonalized. Consider the Bloch sums
\begin{equation}
  \label{eq:14}
  \begin{split}
  \phi^{\pm i}_{\bm k\nu}(\bm r) &= 
  \sum_l e^{i\bm k\cdot(\bm R_l + \bm S_\pm)}\,
  \phi^\pm_i(\bm r - \bm R_l - \bm S_\pm),\\
  & i = 1,\ldots,5,
  \end{split}
\end{equation}
with $\phi^\pm_i$ from (\ref{eq:02}) and form the combinations
\begin{equation}
  \label{eq:15}
  \psi^{s/a,i}_{\bm k\nu} = \frac{1}{\sqrt{2}}
  \bigl(\phi^{+i}_{\bm k\nu} \pm \phi^{-i}_{\bm k\nu}\bigr),
\end{equation}
where $s$ stands for symmetric and $a$ for alternating. Overlap of
neighboring local orbitals is usually neglected in a tb approach. 

\textit{If one, more generally, denotes with $\phi^\pm_i$ Wannier
  functions with the same crystal site symmetry as (\ref{eq:06}), then
  overlap in the original basis (\ref{eq:06}) is automatically included.
  This is done in all what follows. (See also next section.)}

Then, (\ref{eq:15}) immediately yields
\begin{equation}
  \begin{split}
  \label{eq:16}
  H^{ss} &= H^{++} + H^{+-}, \quad H^{aa} = H^{++} - H^{+-}, \\
  H^{sa} &= 0 = H^{as}.
  \end{split}
\end{equation}

The next section deals with the determination of tb parameters from
FPLO8 band structure results. Here, the tb model
(\ref{eq:09}--\ref{eq:16}) for the case of FeSe is analyzed as a typical
example. The corresponding tb parameters are
\begin{equation}
  \label{eq:17}
  \begin{split}
    \epsilon_1 &= \;\;\; 0.014,\\
    \epsilon_2 &= -0.539,\\
    \epsilon_3 &= \;\;\; 0.020,\\
    \epsilon_5 &= -0.581
  \end{split}\quad\;
  \begin{aligned}
    t^{11}_{11}  &= \;\;\; 0.086,   & t^{10}_{16}   &= -0.063,\\
    t^{20}_{11}  &= -0.028,         & t^{21}_{16}   &= \;\;\; 0.017,\\
    t^{11}_{13}  &= -0.056i,        & t^{10}_{18}   &= \;\;\; 0.305i,\\
    t^{11}_{15}  &= -0.109,         & t^{10}_{27}   &= -0.412,\\
    t^{11}_{22}  &= -0.066,         & t^{10}_{29}   &= -0.364i,\\
    t^{11}_{23}  &= \;\;\; 0.089i,  & t^{10}_{2,10} &= \;\;\; 0.338,\\
    t^{11}_{33}  &= \;\;\; 0.232,   & t^{10}_{38}   &= \;\;\; 0.080,\\
    t^{20}_{33}  &= \;\;\; 0.009,   & t^{21}_{38}   &= \;\;\; 0.016,\\
    t^{02}_{33}  &= -0.045          & t^{10}_{49}   &= \;\;\; 0.311,\\
    t^{22}_{33}  &= \;\;\; 0.027,   & t^{21}_{49}   &= -0.019,\\
    t^{11}_{34}  &= \;\;\; 0.099,   & t^{10}_{4,10} &= \;\;\; 0.180i,\\
    t^{11}_{35}  &= \;\;\; 0.146i   & &\textrm{ all in eV}.
  \end{aligned}
\end{equation}
Only in a few cases the on-site energies $\epsilon_i$ and the hopping
matrix elements $t$ directly obtained from the Wannier approach
described in the next section are slightly renormalized in order to
grossly account for the neglected hoppings to higher neighbors.

\begin{figure}[h]
  \centering
  \includegraphics[scale=0.32,angle=-90,clip=]{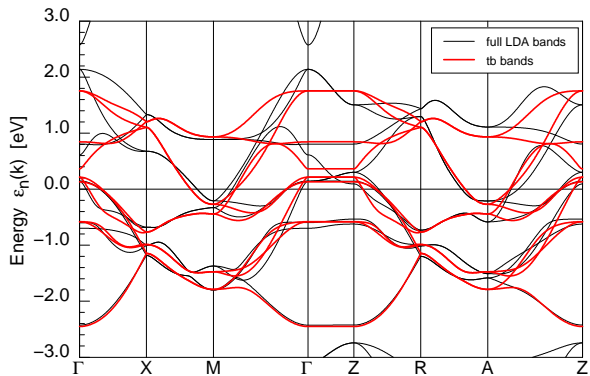}
  \caption{(color online) Comparison of the 2D tb bands with the full 3D
    LDA bands of FeSe.}
  \label{fig:09}
\end{figure}

\begin{figure}[h]
  \centering
  \includegraphics[scale=0.32,angle=-90,clip=]{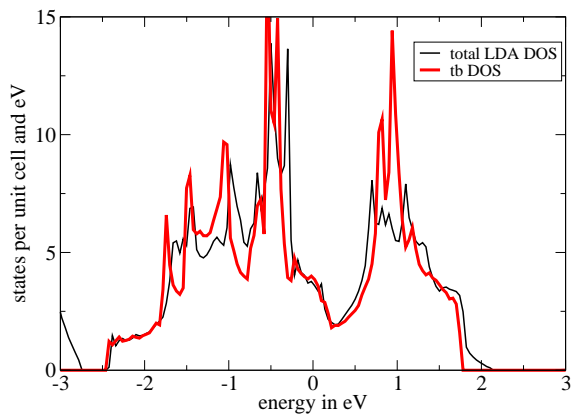}
  \caption{(color online) Comparison of the 2D tb DOS with the full 3D
    LDA DOS of FeSe.}
  \label{fig:10}
\end{figure}

Fig.~\ref{fig:09} shows the obtained 2D two to five neighbor tb bands
in comparison with the full LDA bands of the 3D FeSe crystal. For good
reasons the focus of this treatment is on the vicinity of the Fermi
level.  On the first glance it seems not to be an optimal fit there, in
particular not around the points M and A (see Fig.~\ref{fig:2} for M,
the points Z, R, A are at $k_z=\pi$ above $\Gamma$, X, M). However, due
to hybridization with the Se-$4p$ states the Fe-Fe anti-bonding bands
have a sizable $k_z$-dispersion which cannot be provided by a 2D tb
model. The quality on average of neglect of $k_z$-dispersion of
this tb model is also seen from the comparison of the DOS in
Fig.~\ref{fig:10} showing that the fit is not unreasonable close to
the Fermi level.

\begin{figure}[h]
  \centering
  \includegraphics[scale=0.32,angle=-90,clip=]{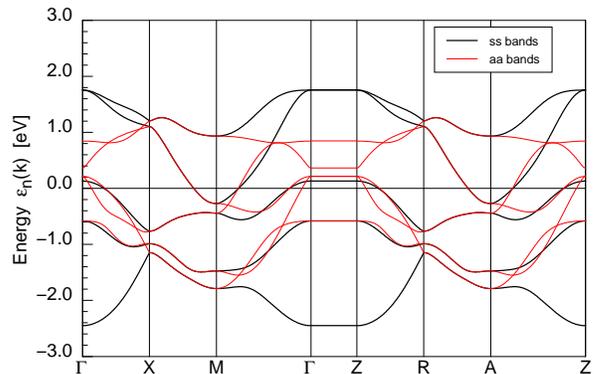}
  \caption{(color online) The ss- and aa-band groups corresponding to
    the Hamiltonian blocks (\ref{eq:14}) for FeSe.}
  \label{fig:11}
\end{figure}

A further feature of the 2D tb model is the block diagonalization
(\ref{eq:16}) of the Hamiltonian matrix by means of lattice symmetry.
There are two five-band groups in this model which do not interact in
the whole BZ. One group comes from symmetric binary Wannier states
(\ref{eq:15}) and one group from alternating ones. From what was said in
the Introduction it follows that each of the two groups contains two
Fe-Fe bonding and two Fe-Fe anti-bonding bands as well as one essentially
non-bonding band. The two groups are shown on Fig.~\ref{fig:11}. Within
each group there is no further decoupling in the whole BZ. In other
words, a further reduction of the number of bands of a tb model
necessarily demands inclusion of (many) higher than second neighbors to
mimic the $\bm k$-dependent interaction with the omitted bands. 

On the side faces of the BZ the two groups are degenerate. As was
demonstrated in Section~II, this degeneracy is a general property of the
P4/nmm space group. It enables to fold out smoothly the ten bands into
five bands in a doubled BZ (Jones-zone \cite{r13}). One may chose the
ss-bands in the BZ proper and the aa-bands folded out, or vice versa. It
is now clearly seen that in any case this is not an out-folding of
anti-bonding bands against bonding bands. Were it so, a pseudo-gap
between bonding and anti-bonding bands could not develop.

The other simplification, the reduction of a $10 \times 10$ Hamiltonian
matrix to a matrix structure with two $5 \times 5$ blocks $H^{++}$
(Hermitian) and $H^{+-}$ (symmetric), is already guaranteed by (4) which
holds for the 122 family too, for which there is no continuous
outfolding. There is no additional benefit besides that. 

In summary so far, without considerably worsening the fit, no less than
the 27 tb parameters (\ref{eq:17}) will do in reasonably modeling the
electronic structure of the iron chalcogenides/pnictides, even if one
focusses on the Fermi level only. This is, unfortunately in stark
contrast to the situation for the cuprates. 

As can be inferred from Fig.~\ref{fig:09}, a 2D tb fit for FeSe cannot be
very satisfactory for quantitative studies even in an energy window of
$\pm 0.1$~eV around the Fermi level, relevant for low temperature
many-body approaches. This is different for the families 1111 and
122. In this respect, although their crystal structures are more
complex their low-energy electronic structure is simpler and really
distinctly 2D.

\begin{figure}[h]
  \centering
\hspace*{0.2mm}\includegraphics[scale=0.315,angle=-90,clip=]{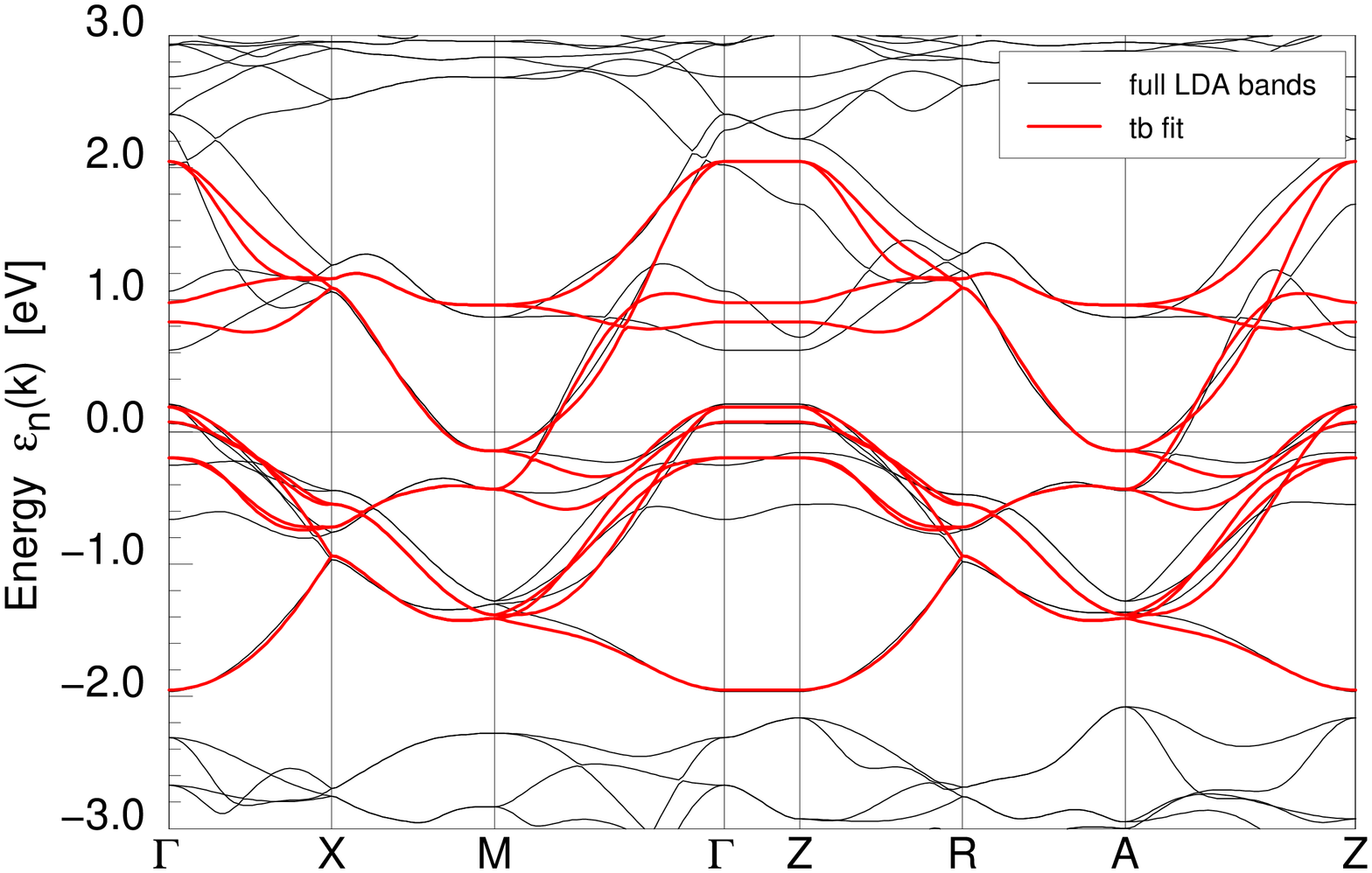}\\
  \includegraphics[scale=0.32,angle=-90,clip=]{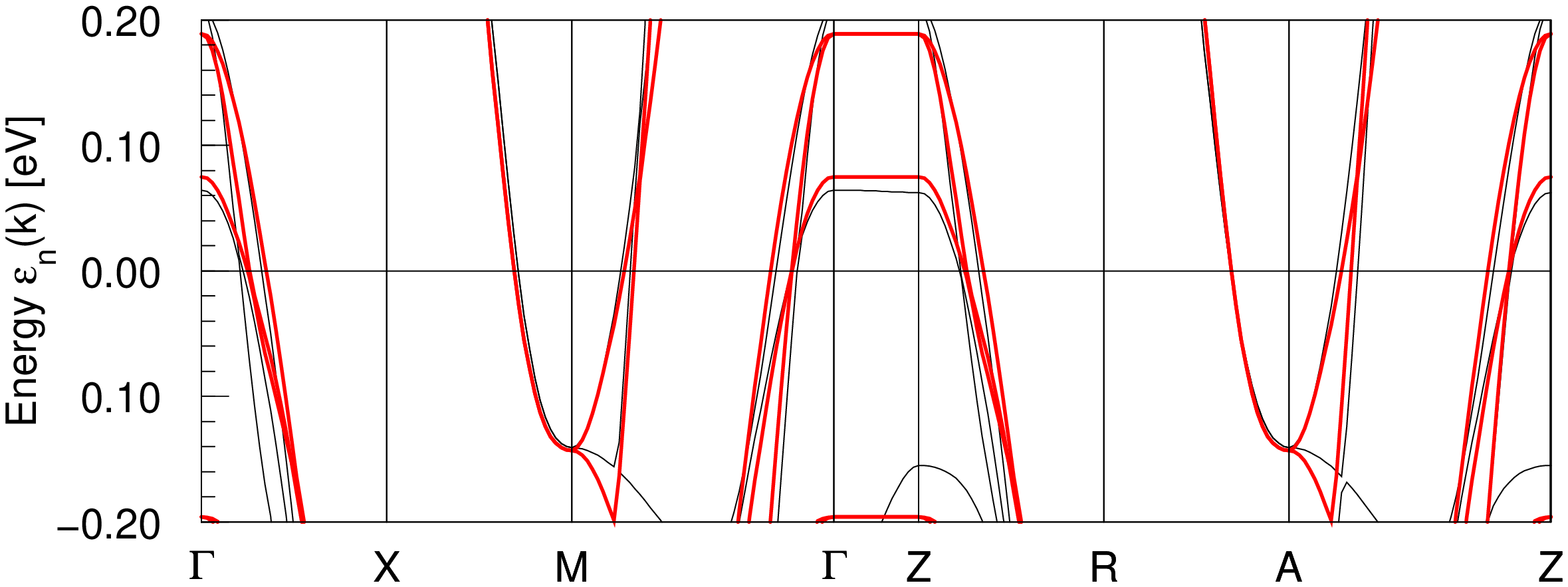}
  \caption{(color online) Comparison of the 2D tb bands with the full 3D
    LDA bands for LaOFeAs. (top) full energy window, (bottom) zoom in the low
    energy region.}
  \label{fig:12}
\end{figure}

For LaOFeAs, the 2D tb parameters are obtained as
\begin{equation}
  \label{eq:19}
  \begin{split}
    \epsilon_1 &= \;\;\; 0.163,\\
    \epsilon_2 &= -0.407,\\
    \epsilon_3 &= \;\;\; 0.053,\\
    \epsilon_5 &= -0.196
  \end{split}\quad\;
  \begin{aligned}
    t^{11}_{11}  &= \;\;\; 0.120,   & t^{10}_{16}   &= -0.167,\\
    t^{20}_{11}  &= -0.029,         & t^{21}_{16}   &= \;\;\; 0.027,\\
    t^{11}_{13}  &= -0.014i,        & t^{10}_{18}   &= \;\;\; 0.224i,\\
    t^{11}_{15}  &= -0.172,         & t^{10}_{27}   &= -0.348,\\
    t^{11}_{22}  &= -0.038,         & t^{10}_{29}   &= -0.315i,\\
    t^{11}_{23}  &= \;\;\; 0.079i,  & t^{10}_{2,10} &= \;\;\; 0.296,\\
    t^{11}_{33}  &= \;\;\; 0.235,   & t^{10}_{38}   &= \;\;\; 0.093,\\
    t^{20}_{33}  &= \;\;\; 0.023,   & t^{21}_{38}   &= \;\;\; 0.026,\\
    t^{02}_{33}  &= -0.025,         & t^{10}_{49}   &= \;\;\; 0.335,\\
    t^{22}_{33}  &= \;\;\; 0.032,   & t^{21}_{49}   &= -0.008,\\
    t^{11}_{34}  &= \;\;\; 0.094,   & t^{10}_{4,10} &= \;\;\; 0.126i,\\
    t^{11}_{35}  &= \;\;\; 0.111i   & &\textrm{ all in eV}.
  \end{aligned}
\end{equation}
The corresponding low-energy 2D bands are compared with the full 3D LDA
bands in Fig.~\ref{fig:12}. Since here the FeAs layers are separated by
LaO layers which have no states in the vicinity of the Fermi level, the
full band structure has a pronounced 2D character there, and
consequently the 2D tb approximation is of high quality. It does
not need improvement by considering 3D dispersion. Nevertheless, for the
LDA or GGA relaxed structure instead of the here considered experimental
Wyckoff positions one of the $z^2$-derived bands crosses the Fermi level
instead of the $xy$-derived one. This is why many published band
structures show a hole pocket around Z instead of the third hole
cylinder.

On the expense of an increase of the number of tb parameters to 44, of
course a better overall fit of the 2D band structure in a larger energy
window can be obtained.\cite{r0}

Somewhat surprisingly, the situation for BaFe$_2$As$_2$ in a narrower
energy window is similar although not equally perfect. This allows to
treat the low-energy band structure of the 122 family within the same
space group P4/nmm (and BZ) which is the space group for one triple
layer of FeAs, although the space group of the full 3D crystal is I4/mmm
(due to an $n$-reflection of adjacently stacked FeAs triple layers). As
a convenient consequence, the Hamiltonian (\ref{eq:09}, \ref{eq:11})
with its simplifying block structure (\ref{eq:13}) applies.

\begin{figure}[h]
  \centering
  \hspace*{0.2mm}\includegraphics[scale=0.315,angle=-90,clip=]{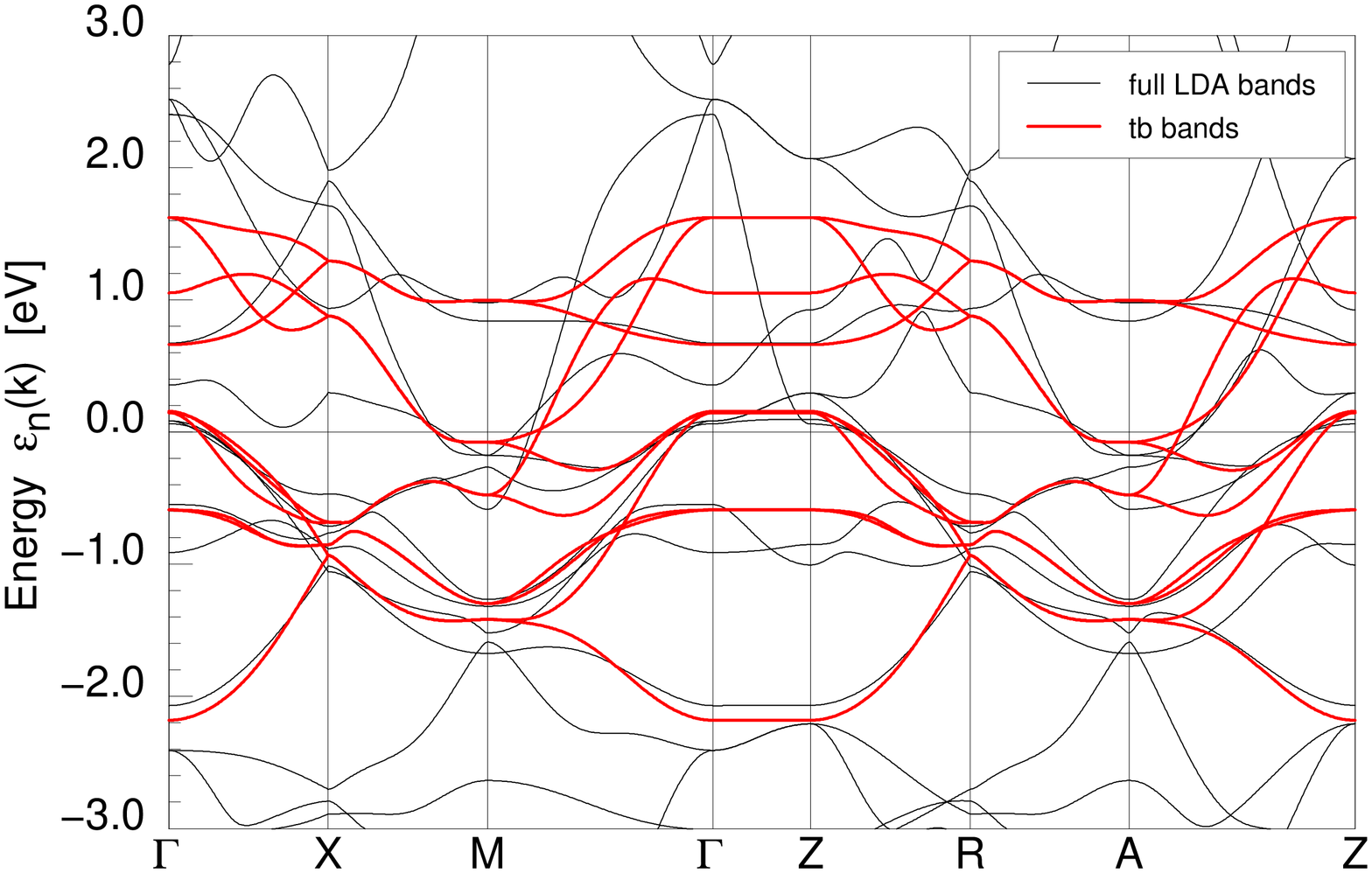}\\
  \includegraphics[scale=0.32,angle=-90,clip=]{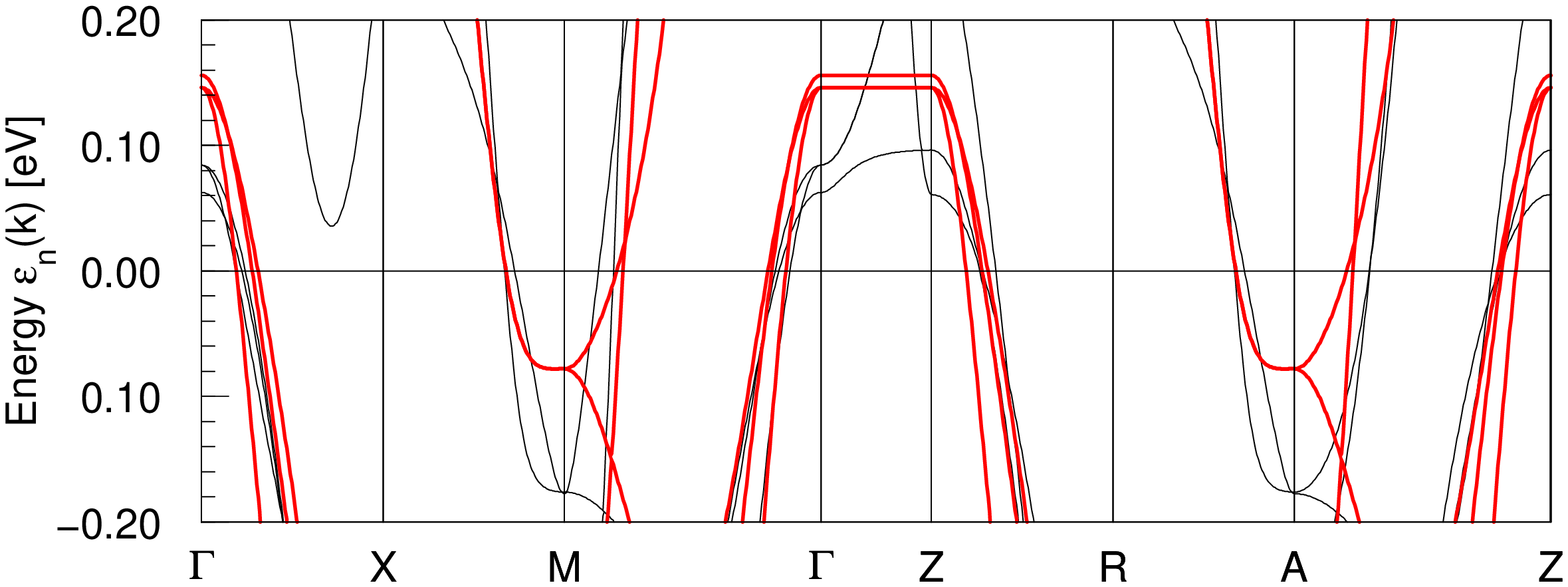}
  \caption{(color online) Like Fig. \ref{fig:11} for BaFe$_2$As$_2$.
    Above the Fermi level the full LDA bands are strongly hybridized
    with Ba $5d$ states.}
  \label{fig:13}
\end{figure}

The 2D tb parameters for BaFe$_2$As$_2$ are obtained as
\begin{equation}
  \label{eq:20}
  \begin{split}
    \epsilon_1 &= \;\;\; 0.172,\\
    \epsilon_2 &= -0.236,\\
    \epsilon_3 &= \;\;\; 0.000,\\
    \epsilon_5 &= -0.590
  \end{split}\quad\;
  \begin{aligned}
    t^{11}_{11}  &= \;\;\; 0.135,   & t^{10}_{16}   &= -0.196,\\
    t^{20}_{11}  &= -0.027,         & t^{21}_{16}   &= \;\;\; 0.042,\\
    t^{11}_{13}  &= -0.024i,        & t^{10}_{18}   &= \;\;\; 0.218i,\\
    t^{11}_{15}  &= -0.131,         & t^{10}_{27}   &= -0.355,\\
    t^{11}_{22}  &= -0.131,         & t^{10}_{29}   &= -0.365i,\\
    t^{11}_{23}  &= \;\;\; 0.103i,  & t^{10}_{2,10} &= \;\;\; 0.265,\\
    t^{11}_{33}  &= \;\;\; 0.204,   & t^{10}_{38}   &= \;\;\; 0.065,\\
    t^{20}_{33}  &= \;\;\; 0.034,   & t^{21}_{38}   &= \;\;\; 0.020,\\
    t^{02}_{33}  &= -0.048,         & t^{10}_{49}   &= \;\;\; 0.312,\\
    t^{22}_{33}  &= \;\;\; 0.024,   & t^{21}_{49}   &= -0.024,\\
    t^{11}_{34}  &= \;\;\; 0.118,   & t^{10}_{4,10} &= \;\;\; 0.080i,\\
    t^{11}_{35}  &= \;\;\; 0.078i   &&\textrm{ all in eV},
  \end{aligned}
\end{equation}
and the low-energy bands together with the full 3D LDA bands are shown
on Fig.~\ref{fig:13}. (Note, that we used the same path along the BZ as
for the other cases, although the symmetry of the 122 system differs.
This, however, facilitates comparision.)
Of course, in both Figs.~\ref{fig:12} and \ref{fig:13} the tb bands on
lines $\Gamma$-X-M-$\Gamma$ and Z-R-A-Z are identical, but the full 3D
LDA bands are not and the plot in particular also shows the distinct 2D
character of the full bands and the quality of the 2D fit.

For later use we also give the 2D tb parameters for LiFeAs valid in
connection with the Hamiltonian (\ref{eq:09}, \ref{eq:11}) although
without a 3D generalization provided in Sec.~V the fit would not even
qualitatively be correct. They are
\begin{equation}
  \label{eq:21}
  \begin{split}
    \epsilon_1 &= -0.188,\\
    \epsilon_2 &= -0.521,\\
    \epsilon_3 &= \;\;\; 0.200,\\
    \epsilon_5 &= -0.609
  \end{split}\quad\;
  \begin{aligned}
    t^{11}_{11}  &= \;\;\; 0.079,   & t^{10}_{16}   &= -0.016,\\
    t^{20}_{11}  &= \;\;\; 0.020,   & t^{21}_{16}   &= \;\;\; 0.013,\\
    t^{11}_{13}  &= -0.090i,        & t^{10}_{18}   &= \;\;\; 0.281i,\\
    t^{11}_{15}  &= -0.060,         & t^{10}_{27}   &= -0.404,\\
    t^{11}_{22}  &= -0.032,         & t^{10}_{29}   &= -0.353i,\\
    t^{11}_{23}  &= \;\;\; 0.087i,  & t^{10}_{2,10} &= \;\;\; 0.313,\\
    t^{11}_{33}  &= \;\;\; 0.275,   & t^{10}_{38}   &= \;\;\; 0.125,\\
    t^{20}_{33}  &= -0.002,         & t^{21}_{38}   &= \;\;\; 0.056,\\
    t^{02}_{33}  &= -0.107,         & t^{10}_{49}   &= \;\;\; 0.359,\\
    t^{22}_{33}  &= \;\;\; 0.012,   & t^{21}_{49}   &= -0.048,\\
    t^{11}_{34}  &= \;\;\; 0.102,   & t^{10}_{4,10} &= \;\;\; 0.190i,\\
    t^{11}_{35}  &= \;\;\; 0.136i   &&\textrm{ all in eV}.
  \end{aligned}
\end{equation}

\section{FPLO8 Wannier functions}

The determination of a Wannier function (WF) basis of states $| {\bm R}
i \rangle$ with function index $i$ at center $\bm S_i$ in unit cell $\bm
R$, though formally being a simple Fourier transform of the Bloch states
$| {\bm k} \nu \rangle$,
\begin{equation}
  \label{eq:WFFourierBloch}
  | {\bm R} i \rangle = \sum_{{\bm k},\nu} 
  | {\bm k} \nu \rangle\,
  U^{{\bm k}}_{\nu i}\,
  \mathrm{e}^{-i \bm k \cdot \bm R},
\end{equation}
becomes an art due to the inherent ambiguity of the choice of the band
index and ${\bm k}$-vector dependent phase factors $U^{{\bm k}}_{\nu
  i}$.

There are basically three goals pursued, when calculating WFs, the first
being to obtain functions, which form a suitable basis for the setup of
model Hamiltonians. This aim usually implies that the resulting WFs
should resemble basis functions, which have a certain chemical
characteristic. The second goal is to reduce the degrees of freedom in
the resulting model by projecting out high energy sectors resulting in
few-band models. The last goal is to obtain model Hamiltonians with as
few as possible parameters, which is essentially the requirement of
maximal localization of the WFs.  Ref.~\cite{marz97} introduced a
general method of calculating maximally localized WFs. Due to the nature
of the problem at hand this approach is rather involved.

In order to fulfill the condition on the chemical characteristics it is
in general desirable to pose symmetry restrictions on the WFs.  In a
local orbital scheme with an optimized chemical basis set the basis
orbitals usually show those characteristics, however, the resulting
basis is non-orthogonal.  Nevertheless, the local basis orbitals can be
used as a starting point for the definition of WFs of a certain
symmetry, which by construction have a high degree of localization.

We follow the ideas sketched in Ref.~\cite{wei02} and introduce a
projection of the Bloch functions $| {\bm k} \nu \rangle$ onto local test
functions $| {\bm S} i \rangle$. The resulting matrix elements
$\langle {\bm k} \nu | {\bm S} i \rangle$ enter the phase factors
$U^{{\bm k}}_{\nu i}$.  They assign weights to the Bloch states entering
the Fourier transform Eq.  (\ref{eq:WFFourierBloch}) such that the bands
with the largest test function character contribute the most. It is very
conceivable that optimized local basis orbitals are among the best
possible choices of test functions. Thus in the simplest case we choose
$| {\bm S} i \rangle$ to be a local orbital at the site ${\bm S}$ with
orbital quantum numbers $i$.  Alternatively, a linear combination of
local orbitals (molecular orbital, MO) centered at the WF center ${\bm
  S}$ could be used, in which case $i$ denotes the characteristics of
the MO.

The choice of the projector singles out all the bands exhibiting a
particular character.  In order to construct few-band models it is often
desirable to additionally project out certain parts of the band complex
with a particular character, say one is interested in WFs describing the
anti-bonding part only. This is achieved by introducing energy windows
described by functions $h_{\bm S i}(\epsilon_{\bm k\nu})$, which are
basically unity in the relevant energy range and zero everywhere else.

All the projectors put together define the Bloch sums corresponding to
the WFs
  \begin{equation}
    \label{eq:blochsumWF}
    | {\bm k} i \} =
    |{\bm k} \nu \rangle \,
    h_{\bm S i}(\epsilon_{\bm k\nu}) \,
    \langle {\bm k} \nu | {\bm S} i \rangle
  \end{equation}
which  in general will be non-orthogonal. A subsequent symmetric
orthogonalization defines the orthogonal WF Bloch sums.
\begin{equation}
  \label{eq:OrthoBlochSumWF}
  | {\bm k} i \rangle =
  \sum_{j}  | {\bm k} j \} \, (S^{-1/2})^{{\bm k}}_{ji}
\end{equation}
where $S^{{\bm k}}_{ji} = \{ {\bm k} j | {\bm k} i \}$. The
resulting phase factors
\begin{equation}
  \label{eq:phasefac}
  U^{{\bm k}}_{\nu i}=
  \sum_j h_{\bm S j}(\epsilon_{\bm k\nu}) \, \langle {\bm k} \nu |
  {\bm S} j \rangle \,
  (S^{-1/2})^{{\bm k}}_{ji} 
\end{equation}
are then used in the Fourier transform Eq. (\ref{eq:WFFourierBloch}) to
calculate the WFs. They have predominately the character of the test
functions $| {\bm S} i \rangle$ and approximate the bands which exhibit
this character and lay in the chosen energy window.  The WFs will
transform according to the symmetry of the test functions.  Due to the
construction the largest part of the WF is actually formed by the test
function itself. Since, we use optimized chemical orbitals as test
functions, the resulting WF will have a high degree of localization.

Furthermore, this construction results in WF band structures which do
not only fit the band energies but also the orbital character of the
bands.  A band, which has predominantly a particular orbital character
in the LDA bands will have the character of the corresponding Wannier
function in the WF bands. Since the WFs are constructed to resemble the
orbitals the Wannier character reflects the original orbital character.
We confirmed this behaviour explicitely by comparing orbital/WF
projected bands.

In the present work, we used the Fe $3d$ orbitals as test functions. The
energy window was chosen such that the c/p-$4p$ bonding bands where
excluded from the energy window and only the 10 Fe-dominated bands were
included.

\section{Three-dimensional dispersion}

As indicated in Sec.~III, there are reasons to consider the
$k_z$-dispersion close to the Fermi level in the family 11, and this
becomes mandatory for the family 111. This increases the number of
parameters considerably since it is connected with hopping through
several layers, and, since there are no vertical bonds, each of these
hoppings adds also one vector $\pm\bm R_1/2$ or $\pm\bm R_2/2$ (for hops
from Fe to c/p) or $\pm\bm T_x$ or $\pm\bm T_y$ (for hops from c/p to
c/p or to a cation).

In the 11 and 111 families, at least three hops are involved from an Fe
layer to the next Fe layer ($t^{rst}$ with $t=\pm 1$), and hence
$t^{rs\pm 1}$ with $(r,s)=(0,0)$ or $(\pm 1,0)$ or $(0,\pm 1)$ or $(\pm
1,\pm 1)$ or $(\pm 2,0)$ or $(0,\pm 2)$ or $(\pm 1,\pm 2)$ or $(\pm
2,\pm 1)$ are of the same order of magnitude and have to be considered
together. (Li or Na orbitals of the 111 family are not involved in the
considered case.) For the vicinity of the Fermi level it suffices to
consider the $k_z$-dispersion of the $xy$, $xz$ and $yz$ bands. The
additions to the Hamiltonian matrix (\ref{eq:09}, \ref{eq:11}) are ($k_z
= \pi$ at BZ-point Z)
\begin{widetext}
\begin{equation}
  \label{eq:50}
  \begin{split}
    H^{++}_{11} &= H^{++}_{11} + \bigl(2t^{001}_{11} +
                   4t^{111}_{11}(\cos k_1 + \cos k_2) +
                   4t^{201}_{11}(\cos k_x + \cos k_y)\bigr)\cos k_z,\\
    H^{++}_{13} &= H^{++}_{13} - 4t^{201}_{14}\sin(2k_y)\sin k_z,\\
    H^{++}_{14} &= H^{++}_{14} - 4t^{201}_{14}\sin(2k_x)\sin k_z,\\
    H^{++}_{33} &= H^{++}_{33} + \bigl(2t^{001}_{33} +
                   4t^{201}_{33}\cos(2k_x) + 4t^{021}_{33}\cos(2k_y)
                   \bigr)\cos k_z,\\
    H^{++}_{44} &= H^{++}_{44} + \bigl(2t^{001}_{33} +
                   4t^{021}_{33}\cos(2k_x) + 4t^{201}_{33}\cos(2k_y)
                   \bigr)\cos k_z,\\[2ex]
    H^{+-}_{16} &= H^{+-}_{16} + 
                   4t^{101}_{16}(\cos k_x + \cos k_y)\cos k_z \;+\\
                &\qquad\quad\;\;  + 2t^{121}_{16}\bigl((\cos(k_1+k_y) +
                   \cos(k_1+k_x))\exp(ik_z) +(\cos(k_2+k_y) +
                   \cos(k_2-k_x)\exp(-ik_z)\bigr),\\
    H^{+-}_{18} &= H^{+-}_{18} - 4\bigl(t^{101}_{18}\sin k_x +
                   t^{101}_{19}\sin k_y\bigr)\sin k_z +
                   2it^{121}_{19}\bigl(
                   \sin(k_1+k_y)\exp(ik_z) -
                   \sin(k_2+k_y)*\exp(-ik_z)\bigr),\\
    H^{+-}_{19} &= H^{+-}_{19} - 4\bigl(t^{101}_{19}\sin k_x +
                   t^{101}_{18}\sin k_y\bigr)\sin k_z +
                   2it^{121}_{19}\bigl(
                   \sin(k_1+k_x)\exp(ik_z) -
                   \sin(k_2-k_x)*\exp(-ik_z)\bigr),\\
    H^{+-}_{38} &= H^{+-}_{38} + 4\bigl(t^{101}_{38}\cos k_x +
                   t^{101}_{49}\cos k_y\bigr)\cos k_z +
                   2t^{121}_{38}\bigl(
                   \cos(k_1+k_x)\exp(ik_z) +
                   \cos(k_2-k_x)\exp(-ik_z)\bigr) \;+\\
                &\hspace*{18.7em} + 2t^{121}_{49}\bigl(
                   \cos(k_1+k_y)\exp(ik_z) +
                   \cos(k_2+k_y)\exp(-ik_z)\bigr),\\
    H^{+-}_{39} &= H^{+-}_{39} + 4it^{101}_{39}\bigl(
                   \cos k_x + \cos k_y\bigr)\sin k_z,\\
    H^{+-}_{49} &= H^{+-}_{49} + 4\bigl(t^{101}_{49}\cos k_x +
                   t^{101}_{38}\cos k_y\bigr)\cos k_z +
                   2t^{121}_{49}\bigl(
                   \cos(k_1+k_x)\exp(ik_z) +
                   \cos(k_2-k_x)\exp(-ik_z)\bigr) \;+\\
                &\hspace*{18.7em} + 2t^{121}_{38}\bigl(
                   \cos(k_1+k_y)\exp(ik_z) +
                   \cos(k_2+k_y)\exp(-ik_z)\bigr).
  \end{split}
\end{equation}
\end{widetext}
Since the Hamiltonian matrix is now complex, (\ref{eq:13}) does not hold
any more, but (\ref{eq:04}, \ref{eq:05}) is still true.

\begin{figure}[h]
  \centering
  \hspace*{0.2mm}\includegraphics[scale=0.315,angle=-90,clip=]{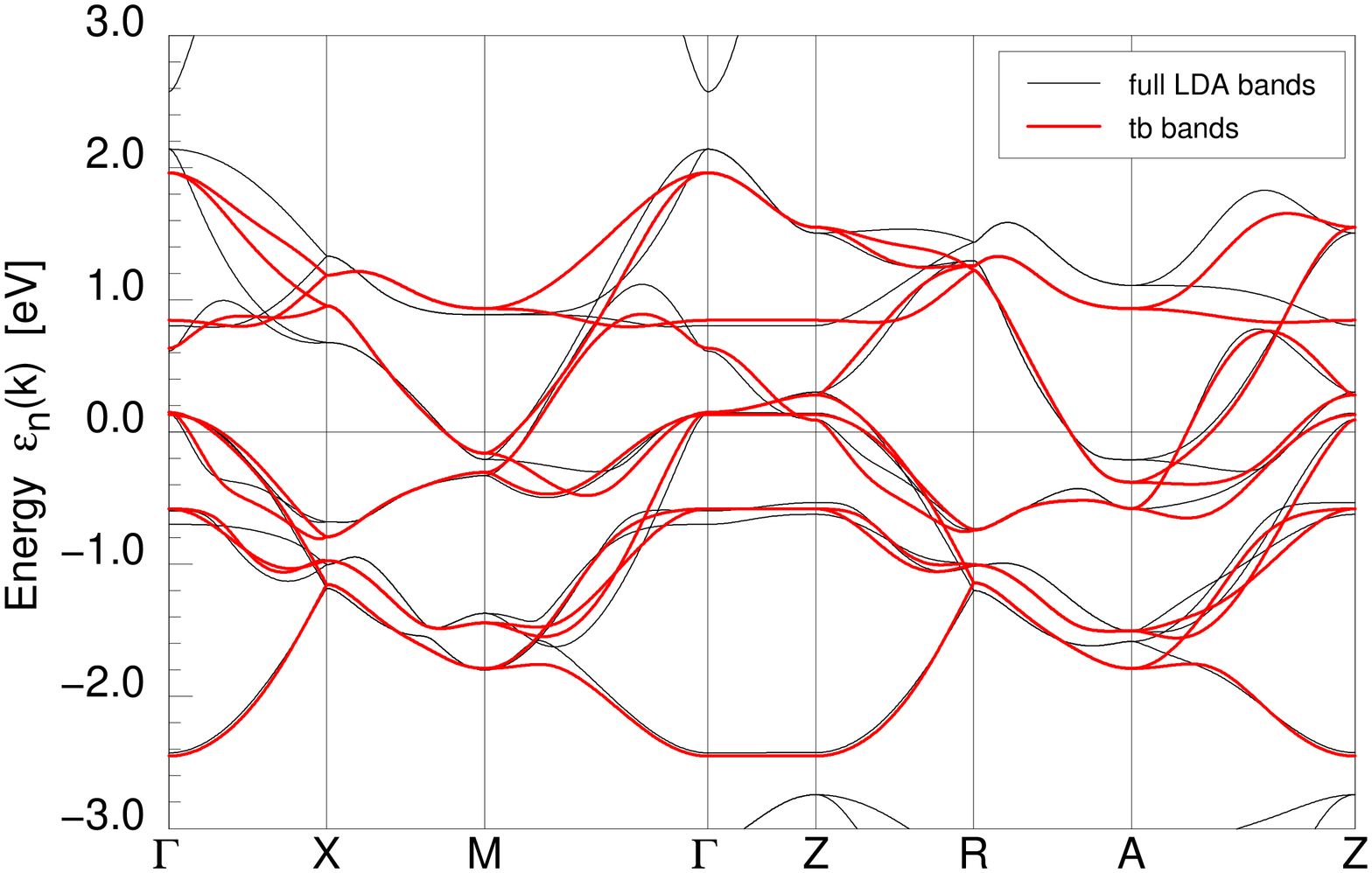}\\
  \includegraphics[scale=0.32,angle=-90,clip=]{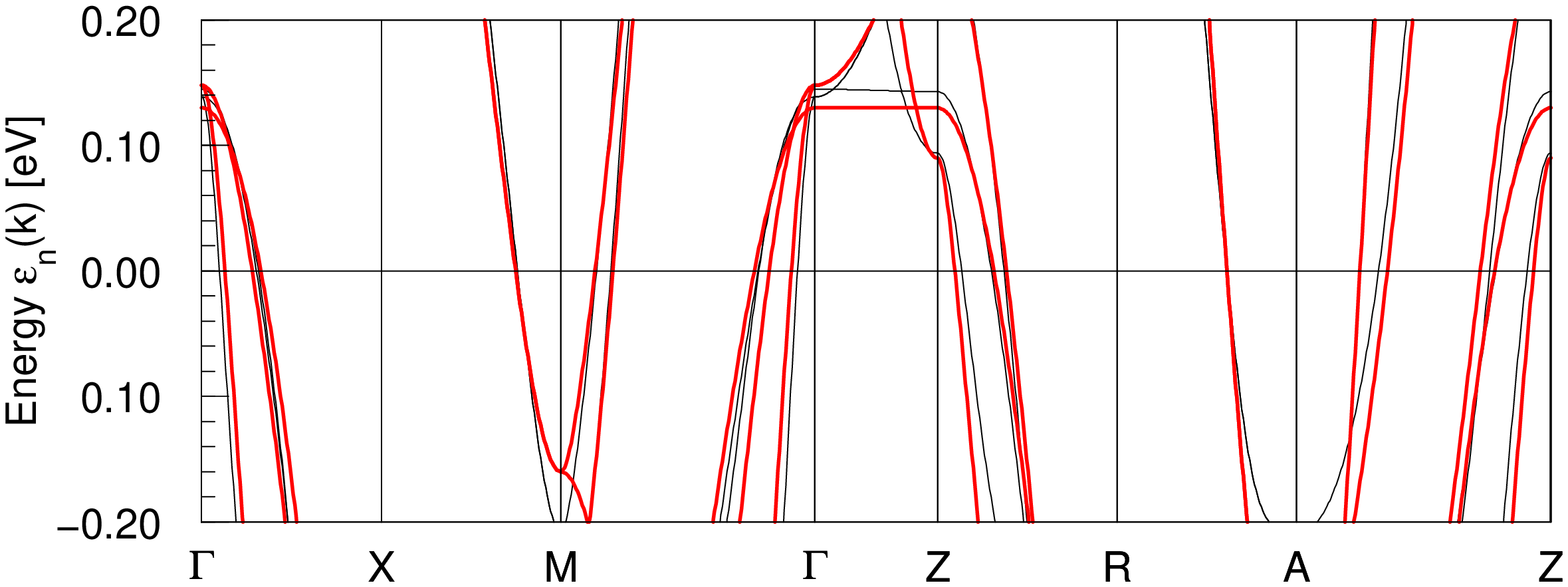}
  \caption{(color online) Comparison of the $k_z$-dispersed tb bands
    with the full LDA bands for FeSe. (top) full energy window, (bottom)
    zoom in the low energy region.}
  \label{fig:20}
\end{figure}

For FeSe, of the 17 new tb parameters contained in (\ref{eq:50}) the 14
non-negligible parameters are (again all in eV)
\begin{equation}
  \label{eq:51}
  \begin{aligned}
    t^{001}_{11}   &= \;\;\; 0,\\
    t^{111}_{11}   &= \;\;\; 0,\\
    t^{201}_{11}   &= \;\;\; 0.017,\\
    t^{201}_{14}   &= \;\;\; 0.030i,\\
    t^{001}_{33}   &= \;\;\; 0.011,\\
    t^{201}_{33}   &= -0.008,\\
    t^{021}_{33}   &= \;\;\; 0.020,
  \end{aligned}\quad\;
  \begin{aligned}
    t^{101}_{16}   &= \;\;\; 0,\\
    t^{211}_{16}   &= -0.017,\\
    t^{101}_{18}   &= \;\;\; 0.009i,\\
    t^{101}_{19}   &= \;\;\; 0.020i,\\
    t^{211}_{19}   &= \;\;\; 0.031i,\\
    t^{101}_{38}   &= \;\;\; 0.006,\\
    t^{211}_{38}   &= -0.003,\\
    t^{101}_{39}   &= \;\;\; 0.015,\\
    t^{101}_{49}   &= \;\;\; 0.025,\\
    t^{211}_{49}   &= \;\;\; 0.006.
  \end{aligned}
\end{equation}
The comparison of the corresponding low-energy tb bands with the full
LDA bands is shown on Fig.~\ref{fig:20}. The fit close to the Fermi
level is probably about the best which can be achieved in this case with
a total of 41 parameters. The 10 Fe-related bands of FeSe are separated
from the rest of the bandstructure by gaps (see Figs.~\ref{fig:01} and
\ref{fig:05}). If one takes the full set of several hundred of tb
parameters calculated from our WFs, the red lines merge nearly perfectly
the black ones in Fig.~\ref{fig:09} for all 10 Fe-3d related bands
shown. This has been checked.

\begin{figure}[h]
  \centering
  \vspace*{2ex}
\hspace*{0.2mm}\includegraphics[scale=0.315,angle=-90,clip=]{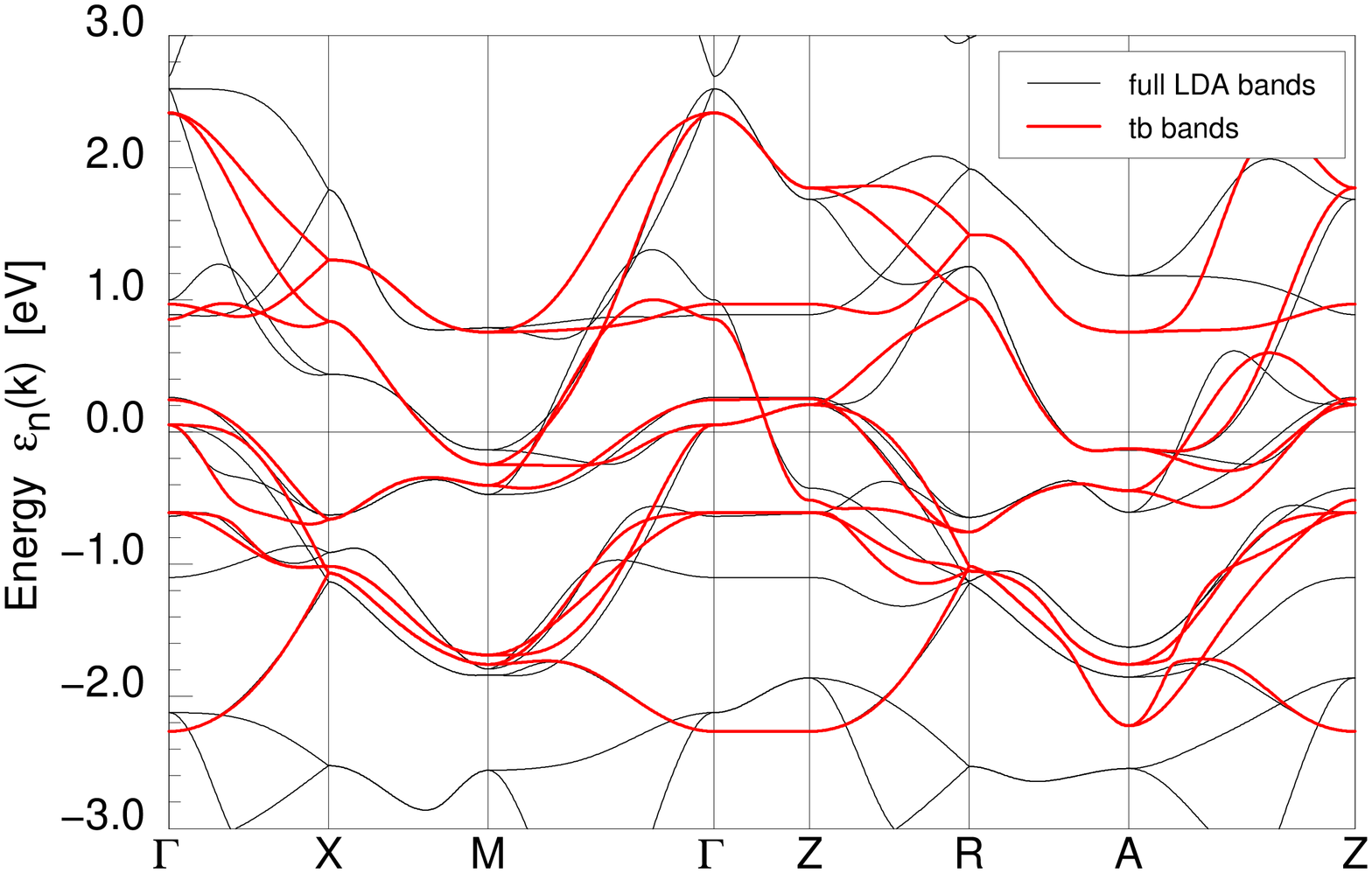}\\
  \includegraphics[scale=0.32,angle=-90,clip=]{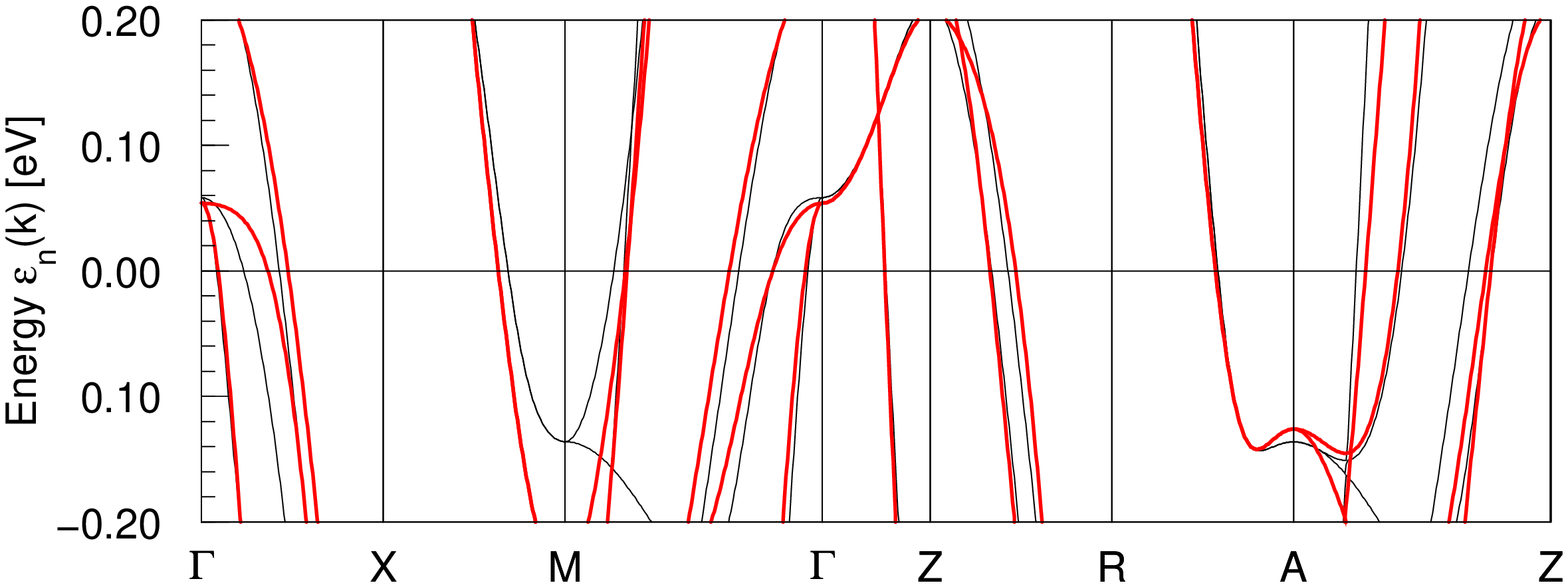}
  \caption{(color online) Like Fig. \ref{fig:20} for LiFeAs.}
  \label{fig:21}
\end{figure}

For LiFeAs, 16 of the 17 new parameters contained in (\ref{eq:50}) are
relevant. They are
\begin{equation}
  \label{eq:52}
  \begin{aligned}
    t^{001}_{11}   &= \;\;\; 0.070,\\
    t^{111}_{11}   &= \;\;\; 0.020,\\
    t^{201}_{11}   &= \;\;\; 0.005,\\
    t^{201}_{14}   &= \;\;\; 0.025i,\\
    t^{001}_{33}   &= -0.004,\\
    t^{201}_{33}   &= -0.003,\\
    t^{021}_{33}   &= \;\;\; 0.031,
  \end{aligned}\quad\;
  \begin{aligned}
    t^{101}_{16}   &= -0.018,\\
    t^{211}_{16}   &= -0.025,\\
    t^{101}_{18}   &= \;\;\; 0.008i,\\
    t^{101}_{19}   &= \;\;\; 0.020i,\\
    t^{211}_{19}   &= \;\;\; 0.022i,\\
    t^{101}_{38}   &= \;\;\; 0,\\
    t^{211}_{38}   &= -0.014,\\
    t^{101}_{39}   &= \;\;\; 0.017,\\
    t^{101}_{49}   &= \;\;\; 0.016,\\
    t^{211}_{49}   &= \;\;\; 0.043.
  \end{aligned}
\end{equation}
The corresponding bands are compared with the full LDA bands in
Fig.~\ref{fig:21}. 

Besides the Fermi radii around M and A now being markedly different,
there now appears a Fermi pocked around $\Gamma$ with a FS crossing even
the line $\Gamma$-Z. One radius around $\Gamma$ could be improved on the
expense of many more hopping parameters over larger distances only,
which is again due to the far extension of the As-$4p$ orbitals.

Let us finally shortly re-examine the 122 system.
With a few $z$-hopping matrix elements, the tb fit in the wide-energy
window shown in Fig. 14 cannot really be improved since the Ba $5d$ states,
strongly hybridizing right above Fermi level, are very
extented. Nevertheless, a few of those hoppings yield a better overall
picture. The matrix elements in addition to (21) are
\begin{equation}
  \label{eq:60}
  \begin{aligned}
    t^{101}_{11}      &= \;\;\; 0.061,\\
    t^{\bar 211}_{14} &= -0.029i,\\
    t^{101}_{33}      &= \;\;\; 0.030,
  \end{aligned}\quad\;
  \begin{aligned}
    t^{001}_{16}      &= -0.161,\\
    t^{001}_{38}      &= \;\;\; 0.080,\\
    t^{11\bar 1}_{16} &= -0.041.
  \end{aligned}
\end{equation}
Due to the different structure the addings to the Hamiltonian matrix are
different from (27). They are
\begin{widetext}
\begin{equation}
  \label{eq:61}
  \begin{split}
    H^{++}_{11} &= H^{++}_{11} + 2t^{101}_{11}(\cos k_x + \cos k_y)
                   \cos k_z,\\
    H^{++}_{13} &= H^{++}_{13} + 2t^{\bar 211}_{14}
                   (i\sin k_x \cos(2k_y) \cos k_z +
                   \cos k_x \sin(2k_y) \sin k_z),\\
    H^{++}_{14} &= H^{++}_{14} + 2t^{\bar 211}_{14}
                   (\sin(2k_x) \cos k_y \sin k_z +
                   i\cos(2k_x) \sin k_y \cos k_z),\\
    H^{++}_{33} &= H^{33}_{33} + 4t^{101}_{33}\cos k_x \cos k_z,\\
    H^{++}_{44} &= H^{33}_{44} + 4t^{101}_{33}\cos k_y \cos k_z,\\[2ex]
    H^{+-}_{16} &= H^{+-}_{16} + 2t^{--1}_{16}\cos k_z +
                   2t^{11\bar 1}_{16}(\cos k_1 \exp(-ik_z) +
                   \cos k_2 \exp(ik_z)),\\
    H^{+-}_{38} &= H^{+-}_{38} + 2t^{001}_{38}\cos k_z,\\
    H^{+-}_{38} &= H^{+-}_{38} + 2t^{001}_{38}\cos k_z.
  \end{split}
\end{equation}
\end{widetext}

\begin{figure}[h]
  \centering
  \includegraphics[scale=0.32,angle=-90,clip=]{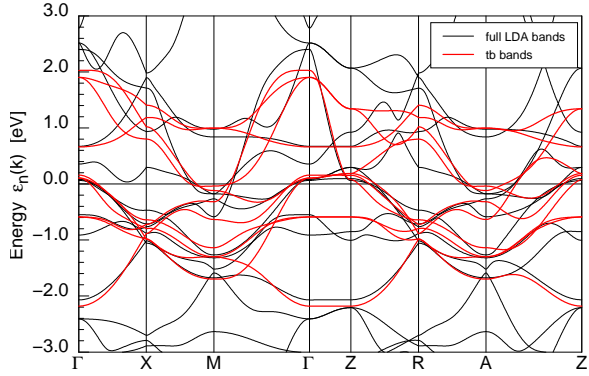}
  \caption{ (color online) Comparison of a $k_z$ dispersed 33 parameter
    tb fit of the Fe $3d$ bands with the full LDA bands for
    BaFe$_2$As$_2$.}
  \label{fig:22}
\end{figure}

This 3D tb fit is shown on Fig. \ref{fig:22}. As is seen, there is one
band (essentially of Ba $5d$ character) comming down close to the Fermi
level which is not represented by the Fe-$3d$ tb fit. At variance  to
the case of LiFeAs, here the 3D dispersing bands are, however, all
unoccupied (and even move away from Fermi level under hole doping).

If one plays with the c/p-Wyckoff parameter or with larger ordered
magnetic moments causing large additional exchange potentials and band
shifts, the 3D behavior and maybe a FS crossing the line $\Gamma$-Z can
reappear even for the 1111 and 122 families. This would put serious
questions on the practicability of the tb approach at all, and one would
probably have to go back to a full numerical WF treatment of the band
structure.\cite{Yanagi08}

\section{Conclusions}

We have demonstrated that the band structures of the non-magnetic
undoped iron based superconductors for all four families have a quite
complex multi-orbital character with all 10 Fe-$3d$ orbitals per unit
cell involved in band states in a 0.1~eV vicinity of the Fermi
level. Thereby, the states at the Fermi level are mainly of $xz$, $yz$ and
$xy$ character, respectively. However, departing from the Fermi level, the
other Fe-$3d$ orbitals and even the c/p-$4p$ orbitals start to hybridize
and largely influence the Fermi velocities.

While the most investigated families 1111 and 122 exhibit clear 2D
behavior in the low electronic energy regime, this is much less the case
for the family 11 and not at all true for the family 111. Minimum
quantitatively correct tb models are presented for FeSe, LiFeAs, LaOFeAs
and BaFe$_2$As$_2$ as representatives for the four families.

A minimum number of 27 tb parameters seems not to be reducible without
severe loss of accuracy of the corresponding Fermi radii, Fermi
velocities and orbital character on the FS. For the families 11 and 111
even that is not sufficient, and the minimum number of tb parameters
goes beyond 40.

\acknowledgements
We benefited from discussions with Igor Mazin, Michelle Johannes and Wei
Ku.

\end{document}